\pdfoutput=1
\documentclass[
pdflatex,sn-mathphys,
Numbered]{sn-jnl}
\usepackage{manyfoot}

\usepackage[T1]{fontenc}
\usepackage[utf8]{inputenc}
\usepackage{amsmath}
\usepackage[title]{appendix}
\usepackage{booktabs}
\usepackage{import}
\usepackage{multirow}
\usepackage{subfig}
\usepackage{graphicx}
\usepackage{grffile}
\usepackage{makecell}

\usepackage{float}
\usepackage{placeins}

\usepackage{./commands}
\usepackage{./symbols}
\usepackage{./colors}

\newcommand{\myTitle}{Improved accuracy of continuum surface flux models for metal additive manufacturing melt pool simulations}

\hypersetup{
	pdfinfo={
		Title={\myTitle},
		Author={Nils Much, Magdalena Schreter-Fleischhacker, Peter Munch, Martin Kronbichler, Wolfgang A.~Wall, Christoph Meier},
		Keywords={continuum surface flux model, multi-phase heat transfer, laser powder bed fusion of metals, melt pool thermo-hydrodynamics, finite element method}
	}
}

\begin{document}

\title{\myTitle}

\author*[1]{\fnm{Nils}\sur{Much}}\email{nils.much@tum.de}\equalcont{Co-First-Authorship}
\author[1]{\fnm{Magdalena}\sur{Schreter-Fleischhacker}}\email{magdalena.schreter@tum.de}\equalcont{Co-First-Authorship}
\author[2,3]{\fnm{Peter}\sur{Munch}}\email{peter.munch@it.uu.se}
\author[3,4]{\fnm{Martin}\sur{Kronbichler}}\email{martin.kronbichler@rub.de}
\author[1]{\fnm{Wolfgang A.}\sur{Wall}}\email{wolfgang.a.wall@tum.de}
\author[1]{\fnm{Christoph}\sur{Meier}}\email{christoph.anton.meier@tum.de}

\affil[1]{\orgdiv{Institute for Computational Mechanics}, \orgname{Technical University of Munich}, \orgaddress{\street{Boltzmannstrasse 15}, \city{Garching}, \postcode{85748}, \country{Germany}}}
\affil[2]{\orgdiv{Department of Information Technology}, \orgname{Uppsala University}, \orgaddress{\street{Box 337}, \city{Uppsala}, \postcode{75105}, \country{Sweden}}}
\affil[3]{\orgdiv{Institute for High-Performance Scientific Computing}, \orgname{University of Augsburg}, \orgaddress{\street{Universitätsstraße 12a}, \city{Augsburg}, \postcode{86159}, \country{Germany}}}
\affil[4]{\orgdiv{Faculty of Mathematics}, \orgname{Ruhr University Bochum}, \orgaddress{\street{Universitätsstraße 150}, \city{Bochum}, \postcode{44780}, \country{Germany}}}

\abstract{
Computational modeling of the melt pool dynamics in laser-based powder bed fusion metal additive manufacturing (\PBFAM) promises to shed light on fundamental mechanisms of defect generation.
These processes are accompanied by rapid evaporation so that the evaporation-induced recoil pressure and cooling arise as major driving forces for fluid dynamics and temperature evolution.
The magnitude of these interface fluxes depends exponentially on the melt pool surface temperature, which, therefore, has to be predicted with high accuracy.
The present work utilizes a diffuse interface finite element model based on a continuum surface flux (CSF) description of interface fluxes to study dimensionally reduced thermal two-phase problems representative for \PBFAM in a finite element framework.
It is demonstrated that the extreme temperature gradients combined with the high ratios of
material properties
between metal and ambient gas lead to significant errors in the interface temperatures and fluxes when classical CSF approaches, along with typical interface thicknesses and discretizations, are applied.
It is expected that this finding is also relevant for other types of diffuse interface \PBFAM melt pool models.
A novel parameter-scaled CSF approach is proposed, which is constructed to yield a smoother temperature field in the diffuse interface region, significantly increasing the solution accuracy.
The interface thickness required to predict the temperature field with a given level of accuracy is less restrictive by at least one order of magnitude for the proposed parameter-scaled approach compared to classical CSF, drastically reducing computational costs.
Finally, we showcase the general applicability of the parameter-scaled CSF to a 3D simulation of stationary laser melting of \PBFAM considering the fully coupled thermo-hydrodynamic multi-phase problem, including phase change.
}

\keywords{
continuum surface flux model,
multi-phase heat transfer,
laser powder bed fusion additive manufacturing,
melt pool thermo-hydrodynamics,
finite element method
}

\maketitle

\section{Introduction}
\label{sec:intro}

Metal additive manufacturing by laser-based powder bed fusion (\PBFAM) is a promising technology that offers unique capabilities for the on-demand production of high-performance metal parts with virtually unlimited design freedom \cite{gibson2015additive}.
In \PBFAM, metal powder is distributed in thin layers on a substrate and selectively molten with a laser beam, forming the so-called melt pool.
A part is built bottom up, and each layer is recoated with a powder layer fused with the subjacent layer.
The resulting stack of layered cross-sections forms the final part.
However, suboptimal process conditions typically lead to quality degrading defects such as porosity, poor surface finish, and residual stresses.
Many of these defects can be attributed to the dynamics in the close vicinity of the melt pool, such as keyhole instabilities, spattering, denudation of the melt track, and balling.
Computational models of the melt pool dynamics promise to shed light on fundamental defect generation mechanisms and to better control part quality.
Additionally, they provide a flexible virtual test environment for new manufacturing approaches without being limited to current manufacturing hardware.

In \PBFAM, the physics in the vicinity of the melt pool constitutes a highly dynamic multi-phase thermo-hydrodynamic problem with phase change.
At the metal-gas interface, there are large jumps in the material properties, with ratios of the density $(\sim 10^{3})$, the specific heat capacity $(\sim 10^{2})$, and the conductivity $(\sim 10^{3})$ between the phases.
The laser-induced heating of the metal substrate induces rapid phase changes, including melting, solidification, and evaporation.
Particularly, evaporation-induced recoil pressure and evaporative cooling lead to strong discontinuities in the heat flux and the flow field at the metal-gas interface and emerge as major driving forces for fluid dynamics and temperature evolution \cite{khairallah2016laser}.
The magnitude of these interface fluxes depends exponentially on the melt pool surface temperature, which also influences other important interface effects such as temperature-dependent surface tension \cite{korner2013fundamental}.
Therefore, to obtain a realistic prediction of the melt pool behavior, the melt pool surface temperature has to be predicted with high accuracy, which is the focus of the present paper.

The computationally demanding thermo-hydrodynamic problem of melt pool dynamics requires highly robust, efficient, and accurate numerical schemes \cite{cook2020simulation}.
Most existing numerical approaches for modeling melt pool dynamics rely on Eulerian frameworks, including discretization by the finite element method (FEM) \cite{andreotta2017finite,courtois2014complete,khairallah2014mesoscopic,leitz2018fundamental,lin2020conservative,queva2020numerical,zhu2021mixed}, the finite difference method \cite{lee2015mesoscopic,pang2011three,tan2013investigation}, the finite volume method \cite{bayat2019multiphysics,chen2020spattering,geiger20093d,panwisawas2017mesoscale}, and the lattice Boltzmann method \cite{ammer2014simulating,zakirov2020predictive}.
Alternatively, Lagrangian meshfree modeling approaches have also been proposed, e.g., based on smoothed particle hydrodynamics \cite{fuchs2022versatile,furstenau2020generating,luthi2023adaptive,russell2018numerical}.

For modeling approaches of multi-phase problems, a distinction is typically made between sharp interface methods and diffuse interface methods \cite{anderson1998diffuse}.
While sharp interface methods fully maintain the discontinuity at the interface, thus enabling a highly accurate representation of the original multi-phase problem \cite{pang2011three,tan2013investigation}, they require complex modifications of the numerical schemes to represent complex topology changes such as breakup and coalescence.
In addition, they typically suffer from stability issues at high ratios of material properties between the phases \cite{schott2014new}.

To overcome the aforementioned issues of sharp interface methods and for a straightforward implementation, diffuse interface methods have been introduced \cite{anderson1998diffuse}.
They have typically been incorporated in existing melt pool models, including Eulerian frameworks \cite{bayat2019multiphysics,courtois2014complete,lee2015mesoscopic,leitz2018fundamental,lin2020conservative,panwisawas2017mesoscale,queva2020numerical,zhu2021mixed,schreter2024consistent} as well as meshfree methods \cite{fuchs2022versatile,luthi2023adaptive,russell2018numerical}.
In addition, the diffuse interface approach is promising for explicitly resolving the evaporation effects \cite{schreter2024consistent}, i.e., the liquid-gas phase transformation as well as the resulting vapor jet and pressure jump, which is typically neglected and thus still pending in the field of melt pool modeling.
In diffuse interface methods, a smooth transition of the properties between the fluids is assumed together with a regularized representation of interface jump conditions over a finite but small thickness of the interface region.
This assumption introduces an inherent modeling error and leads to a less accurate representation of the interface compared to sharp interface methods.
Nevertheless, they are mathematically consistent such that the solution converges to the sharp interface model as the interface thickness decreases and are considered to provide robust solutions.

A popular choice for regularized modeling of interface fluxes in diffuse methods is the classical continuum surface flux (CSF) model according to Brackbill et al.~\cite{brackbill1992continuum}, which was originally introduced to model surface tension effects in two-phase flows but is also employed for other types of interface fluxes, e.g., for heat fluxes in \cite{lin2020conservative}.
However, the usage of classical CSF models for interface fluxes, together with high ratios of the material parameters between the phases, can lead to significant modeling errors.
Let us consider a representative scenario from our intended application to \PBFAM: using classical CSF for modeling laser-induced surface heating in two-phase heat transfer with a high thermal mass ratio between the solid/liquid metal phase and the gas phase can lead to the nonphysical effect that the peak temperature is in the gas phase and not at the interface as expected.
As a result, the interface temperature is mispredicted, which directly affects the accuracy of temperature-dependent interface fluxes, such as the evaporation-induced recoil pressure and, consequently, the predicted dynamics of the melt pool.
However, although the diffuse interface approach is the most popular choice in computational melt pool models, the inherent modeling error and the convergence properties, particularly with respect to the critical interface temperature, have never been quantified so far.

The present work deals with a diffuse interface two-phase model based on a CSF description of interface fluxes embedded in a finite element framework.
In this model, we aim to improve the accuracy of interface quantities, such as interface temperature and temperature-dependent interface fluxes, for application to \PBFAM characterized by high ratios of material properties between phases.
Our particular focus is on achieving an accurate prediction of the thermo-hydrodynamic behavior of the melt pool.
To this end, we specify the objectives as follows:
\begin{itemize}
	\item
	First, we analyze and critically evaluate the accuracy of classical CSF models for interface heat flux modeling in two-phase thermal models.
	To this end, we introduce novel 1D and 2D benchmark examples representative of laser-induced heating in \PBFAM.
	\item
	Second, we propose a parameter-scaled CSF modeling approach.
	The approach is inspired by existing approaches of density-scaled CSF models for surface tension in two-phase flow \cite{kothe1996volume,yokoi2014density}.
	It is constructed primarily for a regularized representation of interface heat fluxes to yield a smoother approximation of the temperature field in the diffuse interface region, which is vital for an accurate and robust numerical model.
	The predicted temperature and recoil pressure field errors are studied based on selected numerical examples with sharp-interface reference solutions.
	\item
	Third, we propose to compute regularized temperature-dependent interface fluxes -- such as evaporative cooling or recoil pressure -- by evaluating the temperature at the interface midplane.
	This aims to enhance the accuracy compared to traditional diffuse methods that use local temperature values to compute regularized interface fluxes.
\end{itemize}
In addition, we demonstrate the general applicability of the parameter-scaled CSF as a proof-of-principle to a 3D simulation of stationary laser melting considering the fully coupled thermo-hydrodynamic problem, including phase change.
To keep the computational cost feasible for this challenging application, we incorporate novel high-performance aspects by using matrix-free operator evaluation and adaptive mesh refinement provided by the \texttt{deal.II} library \cite{arndt2022deal,kronbichler2018fast}.

This article is structured as follows:
Section~\ref{sec:classicalCSF} provides a review and evaluates the classical CSF model based on a novel benchmark example representing a two-phase heat transfer problem of laser-induced heating during \PBFAM with high ratios of the material parameters.
In Section~\ref{sec:parameterScaledCSF}, we present the novel parameter-scaled CSF model and assess its strengths compared to the classical CSF model.
In Section~\ref{sec:onedimEvaporation}, a novel formulation of temperature-dependent continuous surface fluxes is presented and compared with standard approaches for evaporation-induced fluxes.
In Section~\ref{sec:twodimFixedMeltPool}, we introduce a novel benchmark example for computing the heat transfer in a representative melt pool configuration considering the parameter-scaled CSF and a sharp reference solution.
For demonstrating the general applicability of the presented methods to practically relevant problems of \PBFAM, we show a fully coupled thermo-hydrodynamic simulation of melt pool dynamics by employing the parameter-scaled CSF in Section~\ref{sec:applcation_meltpool}.
The results of the previous sections are subject to a discussion of diffuse interface melt pool models in Section~\ref{sec:discussion}, and Section~\ref{sec:conclusions} provides a conclusion.

\section{Review of classical continuum surface flux modeling}
\label{sec:classicalCSF}

Continuum surface flux (CSF) modeling is a popular numerical method to obtain a smoothed representation of singular fluxes at the interface between two phases, aiming at improving the robustness of a finite-element-based multi-phase framework.
By employing a smoothed approximation of the Dirac delta function, sometimes called a regularized Dirac delta function, a continuous flux density is computed, which typically has support only in a finite but small transition region around the interface midplane.
This enables the application of an interface condition in a continuous manner, which is consistent with the concept of the FEM, without the need to reconstruct a discrete surface representation of the sharp interface.
These aspects can be particularly useful for complex, highly dynamically changing interface topologies such as those encountered in melt pool dynamics.

In the following, we briefly summarize the features of a classical continuum surface flux model.
Based on a convergence study on a novel 1D benchmark example for laser-induced heating and a comparison with a sharp interface reference solution, we point out the potential strengths and weaknesses of using the classical CSF approach for melt pool modeling.

\subsection{The classical CSF model}
\label{sec:classicalCSFintro}

\begin{figure}[tbp!]
	\centering
	\includegraphics{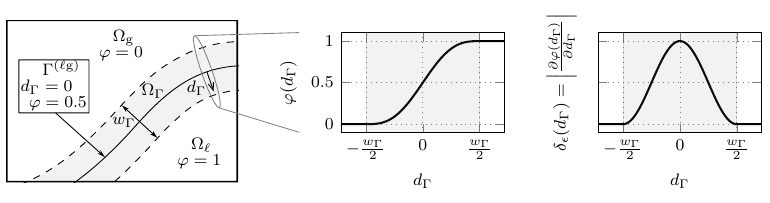}
	\caption{
		Diffuse interface two-phase domain:
		(left) The domain $\Omega$ consists of two phases: a gas domain $\OmegaG$ and a liquid domain $\OmegaL$, that are separated by an interface $\GammaLG$ characterized as the zero contour of the signed distance function $\distance$.
		A narrow band with thickness $\interfaceThickness$ around $\GammaLG$ comprises the diffuse interface region $\DiffuseInterfaceRegion$.
		(center) The indicator $\indicator(\distance)$ specifies the phases, being $\indicator = 0$ in $\OmegaG$ and $\indicator = 1$ in $\OmegaL$ with the transition function according to \eqref{eq:indicator}.
		In $\DiffuseInterfaceRegion$, $\indicator(\distance)$ transitions smoothly between the two phases, attaining $\indicator = 0.5$ at $\GammaLG$.
		(right) Shape of the symmetric delta function $\deltaFunction(\indicator)$ of the classical CSF model, i.e., the norm of the indicator gradient \eqref{eq:norm_of_indicator_gradient}.
	}
	\label{fig:phases_indicator_delta}
\end{figure}

Brackbill et al.~\cite{brackbill1992continuum} proposed one of the first CSF models for representing surface tension of two-phase incompressible flow, which serves as a basis for the present summary and is denoted as \emph{classical CSF}.
We consider a two-phase domain ${\Omega = \OmegaG \cup \OmegaL \cup \DiffuseInterfaceRegion \subset \{\Bx \in \Real^{n}\}}$ with dimensionality $n \in \{1,2,3\}$, that is occupied by a gas phase $\OmegaG$ and a liquid phase $\OmegaL$ as shown in the left panel of Fig.~\ref{fig:phases_indicator_delta}.
For the sake of simplicity, we neglect the fluid dynamics in the first part of this work and use the term liquid phase also for any metal phase that may be molten or solid metal since we assume that the thermal material properties remain constant.
The two phases are separated by a diffuse interface region $\DiffuseInterfaceRegion$, referred to as the interface in the following.
It is defined as a narrow band around the interface midplane $\GammaLG$ with interface thickness $\interfaceThickness > 0$.
The signed distance $\distance(\Bx)$ of a point $\Bx$ represents the distance to the interface midplane $\GammaLG$ being negative in $\OmegaG$ and positive in $\OmegaL$ and defining $\GammaLG$ as its zero contour.
Typically, a smoothed indicator function $\indicator(\distance) \in [0, 1]$ is introduced to distinguish between the two phases, here indicating $\OmegaG$ at $\indicator = 0$ and $\OmegaL$ at $\indicator = 1$, and defining $\GammaLG$ as the \num{0.5} contour of $\indicator(\distance)$.
We choose a sine-based interpolation function according to \cite{sussman1994level, peskin2002immersed}
\begin{align} \label{eq:indicator}
	\indicator(\distance(\Bx)) =
	\begin{cases}
		0 & \text{for} \quad \distance \leq -\frac{\interfaceThickness}{2} \\
		\frac{1}{2} + \frac{\distance}{\interfaceThickness} + \frac{1}{2 \pi} \sin{\left(\frac{2 \pi \distance}{\interfaceThickness}\right)} & \text{for} \quad -\frac{\interfaceThickness}{2} < \distance < \frac{\interfaceThickness}{2} \\
		1 & \text{for} \quad \distance \geq \frac{\interfaceThickness}{2}
	\end{cases}
	\text{,}
\end{align}
depicted in the center panel of Fig.~\ref{fig:phases_indicator_delta}, but any other continuous interpolation function is feasible.
Note that although the continuity of the first derivative is not a prerequisite, it is yet an advantage since it signifies a continuous delta function, which fulfills the first postulate given by Peskin \cite{peskin2002immersed}.
With \eqref{eq:indicator}, a continuous representation of varying thermo-physical properties between the two phases can be obtained, e.g., by computing the arithmetic mean of the phase contributions, which is indicated by the subscript $\arithInterp{(\bullet)}$
\begin{align} \label{eq:parameter_transition_arithmetic}
	\arithInterp{\alpha}(\indicator) = \inGas{\alpha}(1-\indicator) + \inLiquid{\alpha}\indicator
\end{align}
where $\alpha$ represents any property associated with the respective values of the two phases, indicated by the subscripts $\inGas{(\bullet)}$ for the gas phase $\OmegaG$ and $\inLiquid{(\bullet)}$ for the liquid phase $\OmegaL$.
In the CSF model, singular interface fluxes $\interfaceFlux$ on $\GammaLG$ are transformed into a volume flux $\CsfInterfaceFlux$ in $\Omega$, indicated by a tilde $(\tilde{\bullet})$, by using a smoothed approximation of the Dirac delta function $\deltaFunction(\indicator)$ such that
\begin{align}
\label{eq:CSF}
	\int_{\Gamma} \interfaceFlux\,\diffd S= \int_{\Omega} \underbrace{ \deltaFunction(\indicator)\,\interfaceFlux}_{\CsfInterfaceFlux}\,\diffd V
\end{align}
holds.
The delta function $\deltaFunction(\indicator)$ is localized to the diffuse interface region $\DiffuseInterfaceRegion$, i.e., it only has non-zero support for $-\frac{\interfaceThickness}{2} \leq \distance \leq \frac{\interfaceThickness}{2}$.
Thus, the delta function has to satisfy
\begin{align} \label{eq:delta_identity}
	\int_{-\frac{\interfaceThickness}{2}}^{\frac{\interfaceThickness}{2}} \deltaFunction(\indicator(\distance)) \, \diffd \distance = 1 \text{.}
\end{align}
This can be obtained by choosing the Euclidean norm of the gradient of the indicator function
\begin{align} \label{eq:norm_of_indicator_gradient}
	\deltaFunction(\indicator) = || \nabla \indicator ||_{2}
\end{align}
as is typical in classical CSF.
The right panel of Fig.~\ref{fig:phases_indicator_delta} shows $\deltaFunction(\indicator)$ across the interface $\DiffuseInterfaceRegion$.
The CSF model effectively replaces a sharp interface with an interface region with a finite but small thickness.

\subsection{Application of the classical CSF model to interface heat ﬂuxes}
\label{sec:classicalCSFthermal}

In the following, we consider the heat transfer equation in the Eulerian two-phase domain $\Omega$, as depicted in the left panel of Fig.~\ref{fig:phases_indicator_delta}, which reads in its general form as
\begin{align} \label{eq:heat_equation}
	\underbrace{\interp{\left(\rho\cp\right)}}_{\volumetricCapacityEff} \left( \fracPartial{T}{t} + \Bu\cdot\nabla T \right) = \nabla \left(\interp{\conductivity}\,\nabla T\right) + \volumetricHeatSource
	\quad\text{ in }\Omega\times[0,t]
\end{align}
with the temperature $T$, the velocity field $\Bu$, the density $\rho$, the (mass-) specific heat capacity $\cp$, the conductivity $\conductivity$, and the time $t$.
Here, the subscript $\interp{(\bullet)}$ denotes the effective material properties for the two-phase mixture.
Note that the effective material properties $\rhoEff$, $\cpEff$, and $\conductivityEff$ are defined as an interpolation between the two phases according to \eqref{eq:parameter_transition_arithmetic}.
The volume-specific heat capacity $\volumetricCapacity = \rho\cp$ is defined as the product of density and specific heat capacity for all phases, neglecting the temperature-dependent expansion of gases assuming incompressible flow.
The heat transfer equation \eqref{eq:heat_equation} is supplemented by an initial condition:
\begin{align} \label{eq:heat_equation_initial_condition}
	T = T_{0}
	\quad\text{ in }\Omega\times[t = 0]
	\text{.}
\end{align}
Dirichlet and Neumann boundary conditions are imposed according to
\begin{alignat}{2}
	T &= \bar{T}
	\quad&&\text{ on }\Gamma_{\*{D},T}\times[0,t] \label{eq:heat_equation_dirichlet} \\
	-\interp{\conductivity}\,\nabla T \cdot \hat{\Bn}&= 0
	\quad&&\text{ on }\Gamma_{\*{N},T}\times[0,t] \label{eq:heat_equation_neumann}
\end{alignat}
with the outward-pointing unit normal vector $\hat{\Bn}$ at the domain boundary ${\partial\Omega = \Gamma_{\*{D},T} \cup \Gamma_{\*{N},T}}$ with ${\Gamma_{\*{D},T} \cap \Gamma_{\*{N},T} = \emptyset}$.
The midplane of the liquid-gas interface $\GammaLG$ is subject to a prescribed external interface heat flux $\interfaceHeatSource$, which we model using the CSF model as a volumetric heat flux $\volumetricHeatSource = \interfaceHeatSource\,\deltaFunction(\indicator)$ with the norm of the indicator gradient \eqref{eq:norm_of_indicator_gradient}.

\begin{figure}[bp!]
	\centering
	\includegraphics{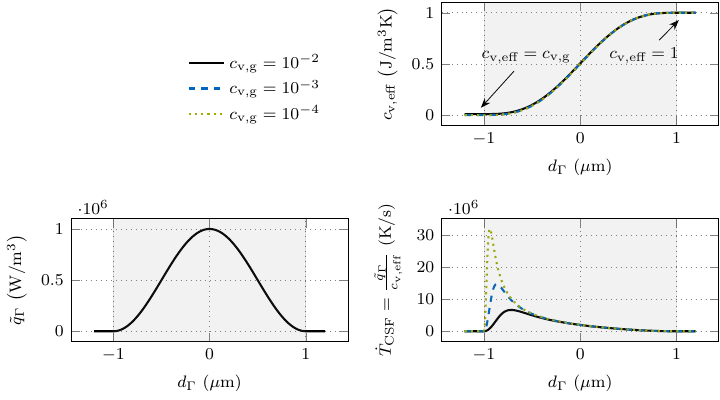}
	\caption{
		Classical CSF modeling of an interface heat flux in the 1D heat equation for different ratios in the volume-specific heat capacity:
		(upper right) effective volume-specific heat capacity $\volumetricCapacityEff = \volumetricCapacityArith$ \eqref{eq:parameter_transition_arithmetic} for the ratios $\frac{\volumetricCapacityLiquid}{\volumetricCapacityGas}\in\{\num{e2}, \num{e3}, \num{e4}\}$ with the fixed value $\volumetricCapacityLiquid = \SI{1}{J\per m^3K}$;
		(lower left) continuum surface heat flux $\volumetricHeatSource = \interfaceHeatSource\,\deltaFunction$ with $\interfaceHeatSource = \SI{1}{W\per m^2}$ and the smoothed approximation of the Dirac delta function $\deltaFunction$ defined by \eqref{eq:norm_of_indicator_gradient};
		(lower right) temperature rate $\dot{T}_{\*{CSF}}$ for the different ratios of the volume-specific heat capacities as the result of the continuum surface heat flux divided by the effective volume-specific heat capacity.
	}
	\label{fig:interface_heat_source_by_capacity_symdelta}
\end{figure}

For the sake of demonstration and without losing generality, we consider the conductive heat transfer in a 1D domain $\Omega = \{x \in \Real\}$ and neglect the convective term in the following, defined by the 1D form of the heat equation \eqref{eq:heat_equation}:
\begin{align} \label{eq:onedim_heat_equation}
	\underbrace{\interp{\left(\rho\cp\right)}}_{\volumetricCapacityEff} \, \fracPartial{T}{t} = \fracPartial{}{x} \left(\interp{\conductivity}\,\fracPartial{T}{x}\right) + \volumetricHeatSource
	\quad\text{ in }\Omega\times[0,t]
	\text{.}
\end{align}
In Fig.~\ref{fig:interface_heat_source_by_capacity_symdelta}, different quantities are illustrated over the signed distance to the interface midplane for an interface thickness of $\interfaceThickness = \SI{2}{\mu m}$.
The top right panel shows the effective volume-specific heat capacity $\volumetricCapacityEff = \volumetricCapacityArith$ \eqref{eq:parameter_transition_arithmetic} for the ratios $\volumetricCapacityLiquid / \volumetricCapacityGas\in\{\num{e2}, \num{e3}, \num{e4}\}$ with the fixed value $\volumetricCapacityLiquid = \SI{1}{J\per(m^3K)}$.
Note that the three curves mostly overlay each other because the different values for $\volumetricCapacityGas$ are almost indistinguishable at the shown scale.
In the bottom left panel, the continuum surface heat flux $\volumetricHeatSource$ is illustrated.

To visualize the effects that arise when modeling strongly localized interface source terms such as the laser heat source in \PBFAM by means of classical CSF approaches, the different contributions in \eqref{eq:onedim_heat_equation} shall briefly be discussed.
For this purpose, we consider the Fourier number $\Fo$, which describes the ratio between conductive heat transfer and heat storage and is defined as
\begin{align} \label{eq:fourier_number}
	\Fo = \frac{\conductivity\,\tau}{\rho\cp\,L^{2}}
\end{align}
with the characteristic time and length scales $\tau$ and $L$.
In \PBFAM, the material is rapidly heated, and temperature rates are typically in the order of \SI{e7}{K\per s}.
The heating process is characterized by short time scales $\tau$, and the Fourier number is typically very small ($\Fo \ll 1$).
Thus, in the initial phase of the heating process, the conductive heat transfer is not significant, and most of the incident laser energy is initially transferred into internal energy of the respective material point, accompanied by a rapid temperature rise according to the left-hand side of \eqref{eq:onedim_heat_equation}.

Accordingly, when neglecting the conductive heat flux in \eqref{eq:onedim_heat_equation}, the initially induced temperature rate $\fracPartial{T}{t}$, which we denote as $\dot{T}_{\*{CSF}}$, can be approximated by the fraction of the continuum surface heat flux $\volumetricHeatSource$ and the volume-specific heat capacity $\volumetricCapacityEff$, as shown in the bottom right panel of Fig.~\ref{fig:interface_heat_source_by_capacity_symdelta}.
It can be seen that a decrease of the volume-specific heat capacity $\cpG$ in the gas phase yields a heavily asymmetric profile of the temperature rate with extreme peak values in the region of low volume-specific heat capacity.
Over time, this irregular shape of the temperature rate profile induces an error in the temperature profile, necessitating the choice of a relatively small interface thickness such that the classical CSF model represents the sharp interface limit with sufficient accuracy.
Moreover, the shown temperature rate profiles across the diffuse interface tend to contain steep gradients, which requires an extremely fine discretization for an accurate representation.

Note that the results within the interface region, such as the steep temperature gradients, are inherent artifacts attributed to the diffuse model and have no explicit physical meaning.
However, the mathematical formulation of the smeared interface flux can influence the results in the interface region.
Accordingly, a carefully constructed formulation of the smeared interface flux can effectively mitigate undesired artifacts in the diffuse interface region.
This can improve numerical robustness and accuracy, enabling the usage of coarser meshes, as elaborated in Section~\ref{sec:parameterScaledCSF}.

\subsection{Benchmark example: laser-induced heating of a static surface}
\label{sec:classicalCSFexample}

\begin{figure}[bp!]
	\centering
	\includegraphics{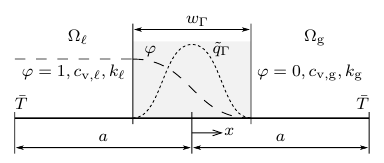}
	\caption{Sketch of the 1D example.}
	\label{fig:onedim_sketch}
\end{figure}

In the following, we propose a simple yet illustrative benchmark example for assessing the strengths and weaknesses of CSF approaches for modeling the laser-induced heat flux at the metal surface, in combination with a typically high ratio of thermal properties between the metal and the inert gas.
To this end, we consider a 1D domain $\Omega = \{x \in [-a, a]\}$ with the length parameter $a = \SI{100}{\mu m}$.
The interface at the center separates the metal phase on the left from the gas phase on the right, as illustrated in Fig.~\ref{fig:onedim_sketch}.
Typical dimensions and process parameters for \PBFAM are employed:
Initially, the temperature is uniform at $T_{0}=\SI{500}{K}$ in the whole domain \eqref{eq:heat_equation_initial_condition}.
At the domain boundary, the temperature is prescribed to $\bar{T}=\SI{500}{K}$ by means of Dirichlet boundary conditions \eqref{eq:heat_equation_dirichlet}.
A surface heat source $\interfaceHeatSource = \SI{e10}{W\per m^2}$ acts upon the interface between the two phases and is regularized using the classical CSF \eqref{eq:CSF}.
The indicator $\indicator$ is prescribed according to \eqref{eq:indicator} with the signed distance of $\distance = -x$ and the interface thickness $\interfaceThickness$.
The material parameters representing \tiSixFour are listed in Table~\ref{tab:param_ti64}.
The resulting ratio of the volume-specific heat capacities is $\volumetricCapacityLiquid/\volumetricCapacityGas = \num{e5}$, and of the conductivities, it is $\inLiquid{\conductivity}/\inGas{\conductivity} = \num{e3}$ between the two phases.
\begin{table}[bp!]
	\caption{Representative material parameters for \tiSixFour.}
	\label{tab:param_ti64}
	\centering
	\begin{tabular}{lcccc}
		\toprule
		Parameter & Symbol & Value & Unit & Reference \\
		\midrule
		conductivity liquid & $\inLiquid{\conductivity}$ & 28.63 & \si{W\per(m\,K)} & \cite{boivineau2006thermophysical} \\
		conductivity gas & $\inGas{\conductivity}$ & 0.02863 & \si{W\per(m\,K)} & \makecell[c]{factor of \num{e3}\\w.r.t.~$\inLiquid{\conductivity}$} \\
		density liquid & $\inLiquid{\rho}$ & 4087 & \si{kg\per m^3} & \cite{mohr2020precise} \\
		density gas & $\inGas{\rho}$ & 4.087 & \si{kg\per m^3} & \makecell[c]{factor of \num{e3}\\w.r.t.~$\inLiquid{\rho}$} \\
		specific heat capacity liquid & $\cpL$ & \num{1130} & \si{J\per(kg\,K)} & \cite{boivineau2006thermophysical} \\
		specific heat capacity gas & $\cpG$ & \num{11.3} & \si{J\per(kg\,K)} & \makecell[c]{factor of \num{e2}\\w.r.t.~$\cpL$} \\
		viscosity liquid & $\inLiquid{\mu}$ & \num{3.5e-3} & \si{kg\per(m\,s)} & \cite{mohr2020precise} \\
		viscosity gas & $\inGas{\mu}$ & \num{3.5e-4} & \si{kg\per(m\,s)} & \makecell[c]{factor of \num{e2}\\w.r.t.~$\inLiquid{\mu}$} \\
		surface tension coefficient & $\sigma$ & \num{1.493} & \si{N\per m} & \cite{mohr2020precise} \\
		laser absorptivity & $\absorptivity$ & \num{0.35} & \dimless & \cite{khairallah2016laser} \\
		boiling temperature & $\Tv$ & 3133 & \si{K} & \cite{zhang2020element} \\
		latent heat of evaporation & $\hv$ & \num{8.84e6} & \si{J\per kg} & calculated using \\
		\makecell[l]{reference temperature for the sum\\of specific enthalpy} & $\Thvref$ & \num{538} & \si{K} & \makecell[c]{the latent heat of \\each component} \\
		molar mass & $M$ & \num{4.78e-2} & \si{kg\per mol} & \cite{lin2023enhanced} \\
		sticking constant & $\cs$ & \num{1} & \dimless & \cite{khairallah2016laser} \\
		liquidus temperature & $\Tl$ & 2200 & \si{K} & numerical value \\
		solidus temperature & $\Ts$ & 1933 & \si{K} & numerical value \\
		\makecell[l]{parameter that depends on the\\mushy zone morphology} & $\darcyMorthology$ & \num{e11} & \si{kg\per(m^3 s)} & numerical value \\
		parameter to avoid division by zero & $\darcyDivZero$ & \num{1} & \dimless & numerical value \\
		\bottomrule
	\end{tabular}
\end{table}
In this work, we focus exclusively on the \PBFAM processes using the material parameters of \tiSixFour.
For this reason, the results are not intended to provide holistic guidelines, as the discretization is likely to be problem-specific.
Issues like non-dimensionalization, therefore, remain topics of future work.
The heat equation \eqref{eq:onedim_heat_equation} is solved using the FEM with evenly spaced finite elements of size $h$ using linear shape functions and implicit Euler time integration with a time step size of $\Delta t = \SI{e-9}{s}$.
In a convergence study, the temporal discretization error was checked to be negligible for that time step size.

In the following, we will consider two different states of the solution: the steady state and the state of an instationary simulation at time $t = \SI{e-5}{s}$.
In the steady state, associated with a vanishing temperature rate in \eqref{eq:onedim_heat_equation}, an analytical solution for the temperature profile is determined considering a sharp representation of the surface heat flux $\interfaceHeatSource$, which reads:
\begin{align} \label{eq:onedim_stat_analytical_max_temperature}
	T(x) = (T_{\*{max}} - \bar{T})\frac{a - |x|}{a} + \bar{T}
	\quad \text{with} \quad
	T_{\*{max}} = \frac{\interfaceHeatSource a}{\kL + \kG} + T_{0}
	\text{.}
\end{align}
The analytical temperature profile resulting from \eqref{eq:onedim_stat_analytical_max_temperature} is used as the reference solution to evaluate the accuracy of the CSF in the steady state.
For the instationary case, the reference solution is determined from a sharp interface solution by applying a surface boundary condition of $q$ at the discrete liquid-gas interface midplane, which aligns with the mesh.
A finite element size of $h = \SI{1.563e-3}{\mu m}$ and a time step size of $\Delta t = \SI{e-10}{s}$ is employed to ensure a converged reference solution.

\begin{figure}[tbp!]
	\centering
	\subfloat[
		Temperature profiles using a CSF interface thickness of $\interfaceThickness = \SI{6}{\mu m}$ for different finite element sizes $h$ at the interface.
	]{
		\includegraphics{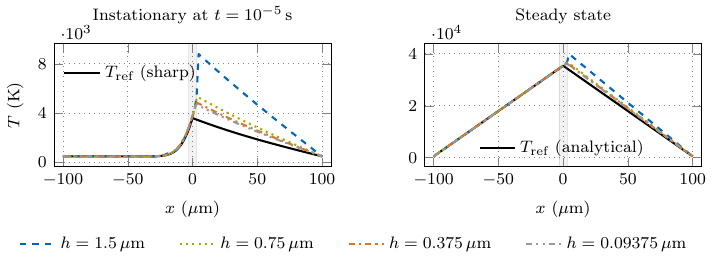}
		\label{fig:onedim_symdelta_dyn_stat_h_stud_consteps_1e-6_T-profiles}
	}
	\\
	\subfloat[
		Relative error in the temperature profile using a CSF interface thickness of $\interfaceThickness = \SI{6}{\mu m}$ for different finite element sizes $h$ at the interface.
		$\numberOfElementsInInterface$ is the number of finite elements across the interface \eqref{eq:numberOfElementsInInterface}.
	]{
		\includegraphics{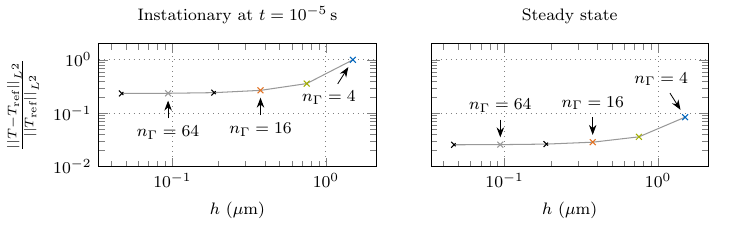}
		\label{fig:onedim_symdelta_dyn_stat_h_stud_consteps_1e-6_T-error}
	}
	\\
	\subfloat[
		Relative error in the temperature profile for different diffuse interface thicknesses $\interfaceThickness$ with ${\numberOfElementsInInterface \in\{8, 16, 32, 64, 128\}}$ finite elements across the interface.
		The interface thickness used in Fig.~\ref{fig:onedim_symdelta_dyn_stat_h_stud_consteps_1e-6_T-error} is annotated.
	]{
		\includegraphics{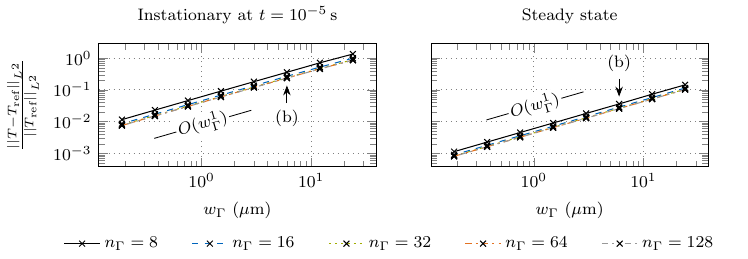}
		\label{fig:onedim_symdelta_dyn_stat_interface_width_stud_ni_128_T-error}
	}
	\caption{Temperature profile resulting from interface heating with an interface heat source of $\interfaceHeatSource = \SI{e10}{W\per m^2}$ using the classical CSF model.
	The instationary reference temperature profile $\Tref$ is determined using a sharp interface approach, and the steady state temperature profile is the analytical solution $\Tref$ according to \eqref{eq:onedim_stat_analytical_max_temperature}.
	}
	\label{fig:onedim_symdelta_studies}
\end{figure}

In Fig.~\ref{fig:onedim_symdelta_studies}, the left panels show the instationary results at $t=\SI{e-5}{s}$ and the right panels the steady state.
Fig.~\ref{fig:onedim_symdelta_dyn_stat_h_stud_consteps_1e-6_T-profiles} shows the temperature profiles of these two scenarios at a constant interface thickness of $\interfaceThickness = \SI{6}{\mu m}$ for different mesh refinements.
The profiles do not converge towards the reference solution because the interface thickness remains constant.
In the liquid domain, the temperature profiles are in good agreement with the reference solution.
However, a significant discrepancy in the temperature profile becomes apparent in the gas domain.
In particular, the peak temperature is overestimated.
Furthermore, the chosen interface thickness is too large, manifested by a significant difference compared to the reference solution, even for the finest mesh.
In the steady state, the discrepancy in the temperature profile becomes smaller since heat is solely transferred by conduction, which is governed by a lower conductivity ratio (\num{e3}) compared to the volume-specific heat capacity (\num{e5}).

Fig.~\ref{fig:onedim_symdelta_dyn_stat_h_stud_consteps_1e-6_T-error} shows the relative temperature error with respect to the sharp interface reference solution for different mesh sizes $h$, including the cases shown in Fig.~\ref{fig:onedim_symdelta_dyn_stat_h_stud_consteps_1e-6_T-profiles}.
Here, the \mbox{$L^{2}$-norm}
\begin{align} \label{eq:L2norm}
	||(\bullet)||_{L^{2}} = \sqrt{\int_{\Omega} (\bullet)^{2}\,\diffd\Omega}
\end{align}
is employed to measure the error.
The resulting number of finite elements across the interface $\numberOfElementsInInterface$ according to
\begin{align} \label{eq:numberOfElementsInInterface}
	\numberOfElementsInInterface = \frac{\interfaceThickness}{h}
\end{align}
is annotated.
Since the interface thickness remains constant, the relative temperature error converges to a non-zero asymptotic value upon mesh refinement.
When doubling $\numberOfElementsInInterface$ from $\numberOfElementsInInterface = 64$ to \num{128}, the change in the relative temperature error falls below \num{1}\%.
Thus, we assume that $\numberOfElementsInInterface = 64$ finite elements across the interface are required to obtain a sufficient mesh resolution for a spatially converged solution.
The overall error in the temperature profile is still significant, which is attributed to the large value of the interface thickness.

In Fig.~\ref{fig:onedim_symdelta_dyn_stat_interface_width_stud_ni_128_T-error}, we show the relative temperature error for different values of the interface thickness and for different discretizations of the interface region.
Reducing the interface thickness while ensuring a sufficient mesh resolution leads to convergence to the sharp interface reference solution.
The relative temperature error decreases with a convergence rate of order $O(\interfaceThickness^{1})$ with respect to the interface thickness.
For $\numberOfElementsInInterface = 64$, the interface thickness has to be less than $\interfaceThickness < \SI{0.248}{\mu m}$ or \num{0.25}\% of the length parameter $a$ for the instationary case and $\interfaceThickness < \SI{2.32}{\mu m}$ or \num{2.32}\% of $a$ for the steady state to attain a tolerance of \num{1}\%.
To put this in perspective, using a homogeneous mesh in the whole 1D domain in Fig.~\ref{fig:onedim_sketch} supporting a sufficient interface thickness with $\numberOfElementsInInterface = 64$ elements across it, we need 51,625 finite elements for the instationary case and 5,510 for the steady state.

Using at least $\numberOfElementsInInterface = 64$ elements over a sufficiently narrow interface is computationally extremely expensive, especially for 3D melt pool simulations.
Hence, when applying classical CSF modeling to melt pool simulations, the high number of finite elements needed for an accurate solution yields a very high computational effort.
This is the motivation for using an alternative formulation for computing continuum surface fluxes, as discussed in the following.

\section{The parameter-scaled continuum surface flux model}
\label{sec:parameterScaledCSF}

To address the poor accuracy and robustness of the classical CSF approach, particularly for high ratios of the material properties between the two phases, the smeared interface flux can be weighted considering the distribution of the material properties.
Regarding surface tension modeling, Kothe et al.~\cite{kothe1996volume} demonstrated that density-scaled CSF, i.e., employing density-scaled delta functions, improves the stability of surface tension force computations in two-phase flow.
In \cite{yokoi2014density}, a density-scaled CSF method is proposed to improve numerical stability and reduce spurious currents due to surface tension forces.
This approach ensures that the magnitude of the surface-tension-induced acceleration is well-balanced across the interface \cite{yokoi2014density}.
It has been used in melt pool simulations to model interface fluxes such as the laser-induced heat source, surface tension, evaporation-induced recoil pressure, and cooling, e.g., by \cite{yan2018fully,zhu2021mixed}.

In the following, we generalize the idea of density-scaled CSF approaches to a parameter-scaled CSF model that accounts for diffuse interface fluxes of various physical problems.
By scaling the continuum surface flux proportional to the parameter that correlates with the inertia of the corresponding equation of the physical problem, we ensure a well-balanced rate of change across the interface.
For the Navier--Stokes momentum equation, the inertia is proportional to the density; thus, the density-scaled CSF is the special case of the parameter-scaled CSF model when applied to interface forces such as the surface tension.
The inertia of the heat equation \eqref{eq:heat_equation} is proportional to the effective volume-specific heat capacity $\volumetricCapacityEff = \interp{\left(\rho\cp\right)}$.
Therefore, we suggest weighting diffuse interface heat fluxes with $\volumetricCapacityEff$.
As will be shown in the numerical study on two-phase heat transfer below, this formulation ensures that the magnitude of the temperature rate is well-balanced across the interface region -- irrespective of the chosen interpolation function for $\volumetricCapacityEff$ between the phases.
In summary, the novelty of the parameter-scaled CSF model lies in its applicability to various physical problems with diffuse interface fluxes.

\subsection{Parameter-scaled delta functions}
\label{sec:parameterScaledCSFdelta}

In the following, we consider the arithmetic mean interpolation $\arithInterp{\alpha}(\indicator)$ according to \eqref{eq:parameter_transition_arithmetic}.
Then, the corresponding parameter-scaled delta function $\deltaSingleWeighted(\indicator)$ is obtained as follows:
\begin{align} \label{eq:delta_singe_weighted}
	\deltaSingleWeighted(\indicator) = \deltaFunction(\indicator) \, \arithInterp{\alpha}(\indicator) \, \deltaSingleWeightedCcorr
	\quad \text{with} \quad
	\deltaSingleWeightedCcorr = \frac{2}{\inGas{\alpha} + \inLiquid{\alpha}}
	\text{.}
\end{align}
Here, $\deltaFunction$ is the initial smoothed Dirac delta function
\eqref{eq:norm_of_indicator_gradient} and the correction factor $\deltaSingleWeightedCcorr$ is chosen such that the parameter-scaled delta function $\deltaSingleWeighted(\indicator)$ satisfies \eqref{eq:delta_identity}.
Suppose $\arithInterp{\alpha}(\indicator)$ represents the density distribution.
In that case, this formulation is identical to a density-scaled CSF of, e.g., \cite{yokoi2014density}, which is beneficial for diffuse interface forces in the momentum equation of the incompressible Navier--Stokes equation.

The choice of the interpolation function is arbitrary with the restriction that it must be a continuous function.
A frequently employed alternative interpolation type to the arithmetic mean is the harmonic mean
\begin{align} \label{eq:parameter_transition_harmonic}
	\harmInterp{\alpha}(\indicator) = \left(\frac{1-\indicator}{\inGas{\alpha}} + \frac{\indicator}{\inLiquid{\alpha}}\right)^{-1}
	\text{,}
\end{align}
where $\harmInterp{\alpha}(\indicator)$ is the interpolated parameter between the phase values $\inGas{\alpha}$ and $\inLiquid{\alpha}$.
The subscript $\harmInterp{(\bullet)}$ refers to the harmonic mean interpolation.
This interpolation type is typically employed in a two-phase flow framework with phase change for the density, e.g., \cite{lee2017direct, schreter2024consistent} and allows the fulfillment of local conservation properties for such models.
Here, the parameter-scaled delta function is chosen according to
\begin{align} \label{eq:delta_harmonic_weighted}
	\deltaHarmonicWeighted(\indicator) &= \deltaFunction(\indicator) \, \harmInterp{\alpha}(\indicator) \, \deltaHarmonicWeightedCcorr \\
	\quad \text{with} \quad
	\deltaHarmonicWeightedCcorr &= \frac{\inGas{\alpha} - \inLiquid{\alpha}}{\inGas{\alpha} \inLiquid{\alpha} \ln{\left(\frac{\inGas{\alpha}}{\inLiquid{\alpha}}\right)}}
	\quad \text{for} \quad
	\inGas{\alpha} > 0 \; \wedge \; \inLiquid{\alpha} > 0 \; \wedge \; \inGas{\alpha} \neq \inLiquid{\alpha}
	\text{,}
	\nonumber
\end{align}
where the correction factor $\deltaHarmonicWeightedCcorr$ is chosen such that the parameter-scaled delta function $\deltaHarmonicWeighted(\indicator)$ satisfies \eqref{eq:delta_identity}.

In the heat equation \eqref{eq:heat_equation}, the thermal mass is proportional to the volume-specific heat capacity $\volumetricCapacity = \rho\cp$.
For CSF interface fluxes in the heat transfer problem, we propose to scale the delta function proportional to the effective volume-specific heat capacity $\volumetricCapacityEff(\indicator)$ to obtain a well-distributed heating rate.
The detailed analysis is shown in Appendix~\ref{app:appendix_CSF_temperature_rate}.
The effective volume-specific heat capacity $\volumetricCapacityEff(\indicator)$ is computed from the product of two parameters, the density $\rhoEff(\indicator)$ and the specific heat capacity $\cpEff(\indicator)$.
In the case that we need to consider the interpolation of two individual parameters in the parameter-scaled delta function, we need to introduce parameter-scaled delta functions with two parameters for both phases.

For two individual parameters that are interpolated with the arithmetic mean interpolation \eqref{eq:parameter_transition_arithmetic}, i.e., $\arithInterp{\alpha}(\indicator)$ and $\arithInterp{\beta}(\indicator)$, we propose the parameter-scaled delta-function
\begin{align} \label{eq:delta_double_weighted}
	\deltaDoubleWeighted(\indicator) &= \deltaFunction(\indicator) \,\arithInterp{\alpha}(\indicator) \, \arithInterp{\beta}(\indicator) \, \deltaDoubleWeightedCcorr \\
	\quad \text{with} \quad
	\deltaDoubleWeightedCcorr &= \frac{6}{2\inGas{\alpha}\inGas{\beta} + \inGas{\alpha}\inLiquid{\beta} + \inLiquid{\alpha}\inGas{\beta} + 2\inLiquid{\alpha}\inLiquid{\beta}}
	\text{,}
	\nonumber
\end{align}
where the correction factor $\deltaDoubleWeightedCcorr$ is computed such that the parameter-scaled delta function $\deltaDoubleWeighted(\indicator)$ satisfies \eqref{eq:delta_identity}.

In the case that one parameter is interpolated using the harmonic mean \eqref{eq:parameter_transition_harmonic}, the following parameter-scaled delta function is used:
\begin{align} \label{eq:delta_harmonic_times_arithmetic_weighted}
	\deltaHarmonicTimesArithWeighted(\indicator) &= \deltaFunction(\indicator) \, \harmInterp{\alpha}(\indicator) \, \arithInterp{\beta}(\indicator) \, \deltaHarmonicTimesArithWeightedCcorr
	\\
	\text{with} \quad
	\deltaHarmonicTimesArithWeightedCcorr &= \left(\inGas{\beta} \, \frac{\inGas{\alpha} \, \inLiquid{\alpha} \, \ln{\left(\frac{\inGas{\alpha}}{\inLiquid{\alpha}}\right)}}{\inGas{\alpha} - \inLiquid{\alpha}} + (\inLiquid{\beta} - \inGas{\beta}) \, \frac{\frac{1}{\inGas{\alpha}} \, \left( \ln{\left(\frac{\inLiquid{\alpha}}{\inGas{\alpha}}\right)} - 1 \right) + \frac{1}{\inLiquid{\alpha}}}{\left( \frac{1}{\inLiquid{\alpha}} - \frac{1}{\inGas{\alpha}} \right)^2} \right)^{-1}
	\nonumber
	\\
	\text{for} \quad \inGas{\alpha} > \,&0 \; \wedge \; \inLiquid{\alpha} > 0 \; \wedge \; \inGas{\alpha} \neq \inLiquid{\alpha}
	\text{.}
	\nonumber
\end{align}
Here, $\alpha$ is interpolated according to \eqref{eq:parameter_transition_harmonic}, $\beta$ is interpolated according to \eqref{eq:parameter_transition_arithmetic} and the correction factor $\deltaHarmonicTimesArithWeightedCcorr$ is chosen such that the parameter-scaled delta function $\deltaHarmonicTimesArithWeighted(\indicator)$ satisfies \eqref{eq:delta_identity}.

\subsection{Application of the parameter-scaled CSF model to interface heat fluxes}
\label{sec:parameterScaledCSFthermal}

To demonstrate the effect of the parameter-scaled CSF, we consider the heat equation \eqref{eq:onedim_heat_equation} on a 1D domain with two phases separated by an interface at the origin.
At the interface, a heat flux $\interfaceHeatSource$ is applied using the parameter-scaled CSF as the volumetric heat flux $\volumetricHeatSource = \interfaceHeatSource\,\deltaI$, where $\deltaI$ is the appropriate delta function for the chosen interpolation of the effective volume-specific heat capacity $\volumetricCapacityEff$.
Table~\ref{tab:parameter_scaled_CSF_cases} lists the four considered interpolation types (V1 - V4) for the interpolation type and delta function pairs.

\begin{table}[bp!]
	\caption{Considered volume-specific heat capacity interpolation types and corresponding parameter-scaled delta functions $\deltaI$ with their scaling parameters.}
	\label{tab:parameter_scaled_CSF_cases}
	\centering
	\begin{tabular}{r|p{1.6cm}p{1.6cm}p{1.6cm}p{2.3cm}l}
		& V1 & V2 & V3 & V4 \\
		\hline
		$\volumetricCapacityEff$ & $\volumetricCapacityArith$ \eqref{eq:parameter_transition_arithmetic} & $\volumetricCapacityHarm$ \eqref{eq:parameter_transition_harmonic} & $\arithInterp{\rho} \cdot c_{\*{p,a}}$ \eqref{eq:parameter_transition_arithmetic} & $\harmInterp{\rho} \cdot c_{\*{p,a}}$ \eqref{eq:parameter_transition_arithmetic}, \eqref{eq:parameter_transition_harmonic} \\
		$\deltaI$ & $\deltaSingleWeighted$ \eqref{eq:delta_singe_weighted} & $\deltaHarmonicWeighted$ \eqref{eq:delta_harmonic_weighted} & $\deltaDoubleWeighted$ \eqref{eq:delta_double_weighted} & $\deltaHarmonicTimesArithWeighted$ \eqref{eq:delta_harmonic_times_arithmetic_weighted} \\
		with & $\alpha_{j} = \volumetricCapacityJ$ & $\alpha_{j} = \volumetricCapacityJ$ & $\alpha_{j} = \rho_{j}$, & $\alpha_{j} = \rho_{j}$, \\
		& & & $\beta_{j} = \cpI$ & $\beta = \cpI$ & for $j\in\{\*{g}, \ell\}$
	\end{tabular}
\end{table}

\begin{figure}[tbp!]
	\centering
	\includegraphics{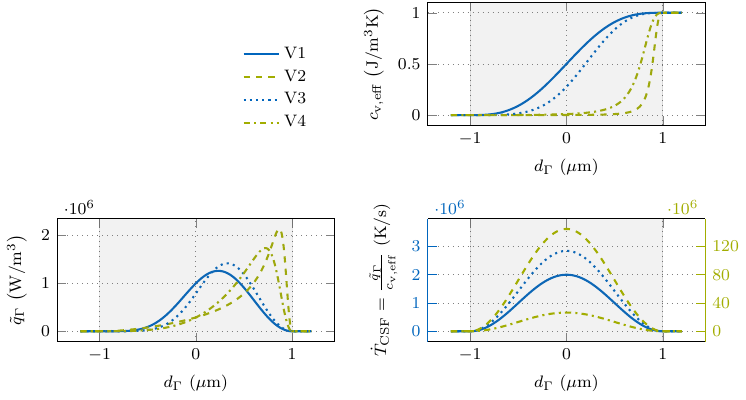}
	\caption{
		Parameter-scaled CSF modeling of an interface heat flux in the 1D heat transfer equation \eqref{eq:onedim_heat_equation} for different interpolations of the volume-specific heat capacity and corresponding delta functions $\deltaI$ for $i\in\{\text{a}; \text{h}; \text{a,a};\text{h,a}\}$ according to Table~\ref{tab:parameter_scaled_CSF_cases}.
		For all cases, the densities of the phases are ${\rhoL = \SI{1}{kg\per m^3}}$ and ${\rhoG = \SI{e-2}{kg\per m^3}}$ and the specific heat capacities are ${\cpL = \SI{1}{J\per(kg\,K)}}$ and ${\cpG = \SI{e-1}{J\per(kg\,K)}}$ which results in the volume-specific heat capacities of ${\volumetricCapacityLiquid = \SI{1}{J\per(m^3 kg)}}$ and ${\volumetricCapacityGas = \SI{e-3}{J\per(m^3 kg)}}$:
		(upper right) effective volume-specific heat capacity $\volumetricCapacity$;
		(lower left) continuum surface heat flux $\volumetricHeatSource = \interfaceHeatSource\,\deltaI$ with $\interfaceHeatSource = \SI{1}{W\per m^2}$ and a smoothed Dirac delta function $\deltaI$;
		(lower right) temperature rate $\dot{T}_{\*{CSF}}$ due to the continuum surface heat flux as the result of the continuum surface heat flux divided by the effective volume-specific heat capacity.
		Due to the significant scale difference, V1 and V3 use the scale on the left $y$-axis, and V2 and V4 use the scale on the right $y$-axis to improve readability.
	}
	\label{fig:interface_heat_source_by_capacity_asymdelta}
\end{figure}

The bottom left panel of Fig.~\ref{fig:interface_heat_source_by_capacity_asymdelta} shows the resulting continuum surface heat flux $\volumetricHeatSource$ over the signed distance to the interface midplane, and the top right panel of Fig.~\ref{fig:interface_heat_source_by_capacity_asymdelta} shows the effective volume-specific heat capacity $\volumetricCapacityEff$ over the diffuse interface, for the different interpolation types listed in Table~\ref{tab:parameter_scaled_CSF_cases}.
The resulting temperature rate $\dot{T}_{\*{CSF}}$ (Fig.~\ref{fig:interface_heat_source_by_capacity_asymdelta} bottom right) has a smooth shape without steep gradients for all cases, compared to the classical CSF in Fig.~\ref{fig:interface_heat_source_by_capacity_symdelta}, which is beneficial for discretization with the FEM.
The shape of the temperature rate profile is independent of the interpolation type and the parameter ratio between the phases, and it follows the shape of the norm of the indicator gradient \eqref{eq:norm_of_indicator_gradient} multiplied by a constant scaling factor.
This is because the parameter-scaled delta functions are designed to yield this result, which is discussed in detail in Appendix~\ref{app:appendix_CSF_temperature_rate} and results in the fact that the appropriate parameter-scaled delta function cancels the interpolation function from the volume-specific heat capacity.

However, the magnitudes of the temperature rates $\dot{T}_{\*{CSF}}$ differ between the cases, necessitating the usage of a different $y$-axis scaled by a factor of 40 for cases V2 and V4, which involve the harmonic mean interpolation \eqref{eq:parameter_transition_harmonic}.
The interpolation of parameters across the interface thickness involves a modeling error attributed to the diffuse interface model.
Any interpolation type is valid as long as the modeling error vanishes within the limit of a small interface thickness, making the diffuse interface model mathematically consistent.
However, the different interpolation types yield different local and average heat capacities for finite interface thicknesses, impacting the physical behavior in the interface region.
The difference in the magnitude of the temperature rate $\dot{T}_{\*{CSF}}$ is due to the reciprocally proportional effective volume-specific heat capacity $\volumetricCapacityEff$.
For the four cases V1-V4, the effective volume-specific heat capacity $\volumetricCapacityEff$ differs significantly within the interface, as seen in the top right panel of Fig.~\ref{fig:interface_heat_source_by_capacity_asymdelta}.
The discrepancies between the four cases only occur in the diffuse interface region $\DiffuseInterfaceRegion$, and they vanish in the limit of small interface thicknesses.
Although there is a difference in the temperature rate $\dot{T}_{\*{CSF}}$ for different interpolations of the effective volume-specific heat capacity $\volumetricCapacityEff$, the spatially integrated values of both, the external heat flux $\interfaceHeatSource$ and the internal energy rate $\rhoCpEff\fracPartial{T}{t}$, remain constant, as is shown in Appendix~\ref{app:appendix_CSF_energy_rate}.
The magnitude of the temperature rate $\dot{T}_{\*{CSF}}$ can serve as an indicator for the performance of the approach.
The following convergence studies suggest that approaches with a low magnitude of the temperature rate profile show the best performance.

\subsection{Investigation of the parameter-scaled CSF model on the laser-induced heating benchmark example}
\label{sec:parameterScaledCSFexample}

\begin{figure}[tbp!]
	\centering
	\subfloat[
	Instationary temperature profiles at $t=\SI{e-5}{s}$ using a CSF interface thickness of $\interfaceThickness = \SI{6}{\mu m}$ for different finite element sizes $h$ at the interface.
	]{
		\includegraphics{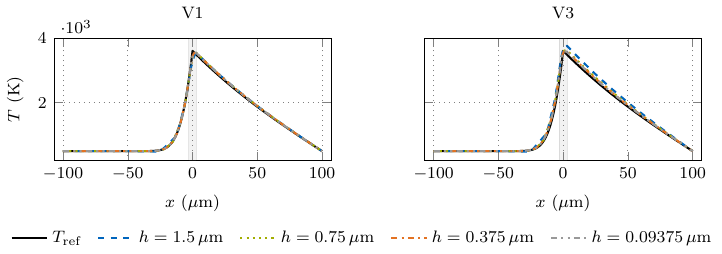}
		\label{fig:onedim_asymdelta_dyn_stat_h_stud_consteps_1e-6_T-profiles}
	}
	\\
	\subfloat[
	Relative error in the instationary temperature profile at $t=\SI{e-5}{s}$ using a CSF interface thickness of $\interfaceThickness = \SI{6}{\mu m}$ for different finite element sizes $h$ at the interface.
	$\numberOfElementsInInterface$ is the number of finite elements across the interface \eqref{eq:numberOfElementsInInterface}.
	]{
		\includegraphics{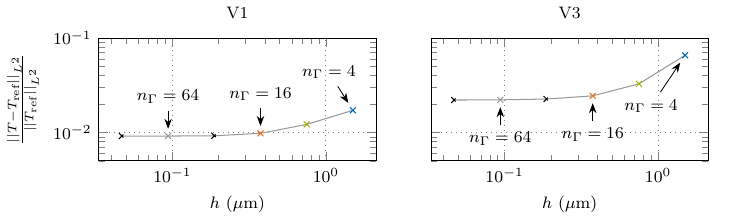}
		\label{fig:onedim_asymdelta_dyn_stat_h_stud_consteps_1e-6_T-error}
	}
	\\
	\subfloat[
	Relative error in the instationary temperature profile at $t=\SI{e-5}{s}$ for different diffuse interface thicknesses $\interfaceThickness$ with ${\numberOfElementsInInterface \in\{8, 16, 32, 64, 128\}}$ finite elements across the interface.
	The interface thickness used in Fig.~\ref{fig:onedim_asymdelta_dyn_stat_h_stud_consteps_1e-6_T-error} is annotated.
	]{
		\includegraphics{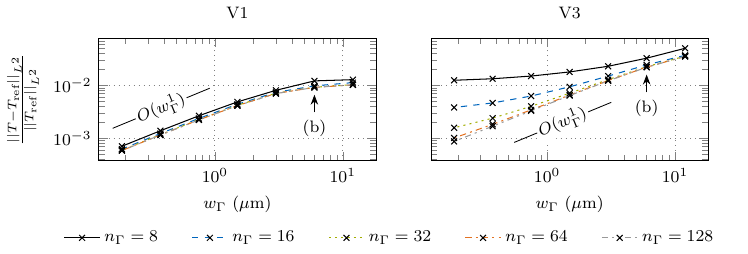}
		\label{fig:onedim_asymdelta_dyn_stat_interface_width_stud_ni_128_T-error}
	}
	\caption{
		Temperature profile resulting from interface heating with an interface heat source of $\interfaceHeatSource = \SI{e10}{W\per m^2}$ using the parameter-scaled CSF approaches V1 and V3, described in Table~\ref{tab:parameter_scaled_CSF_cases}.
		The reference temperature profile $\Tref$ is determined using a sharp interface approach.
	}
	\label{fig:onedim_asymdelta_studies}
\end{figure}

For evaluating the performance of the parameter-scaled CSF for modeling interface heat fluxes, we reconsider the 1D benchmark example illustrated in Fig.~\ref{fig:onedim_sketch} and described in Section~\ref{sec:classicalCSFexample}.
Since the instationary case showed worse accuracy than the steady-state for the original smoothed Dirac delta function, we focus only on the instationary case from this point on.
Fig.~\ref{fig:onedim_asymdelta_studies} shows the instationary results at $t = \SI{e-5}{s}$ for two cases listed in Table~\ref{tab:parameter_scaled_CSF_cases}, considering the interpolation type via an arithmetic mean \eqref{eq:parameter_transition_arithmetic} with the case V1 in the left column and V3 in the right column.

The cases V2 and V4, which involve the harmonic mean interpolation \eqref{eq:parameter_transition_harmonic}, are not discussed in detail in the following.
They have been tested in the same fashion as discussed below but performed not as well as V1 and V3, requiring finer mesh resolution and narrower interface thicknesses to achieve the same accuracy, see Appendix~\ref{app:parameterScaledCSFexampleHarmonic}.
This is attributed to the increase of gradients in the interface region for the relevant measures shown in Fig.~\ref{fig:interface_heat_source_by_capacity_asymdelta}.

In Fig.~\ref{fig:onedim_asymdelta_dyn_stat_h_stud_consteps_1e-6_T-profiles}, the temperature profiles over the domain are shown at a constant interface thickness of $\interfaceThickness = \SI{6}{\mu m}$ for different mesh refinements.
For V1, the temperature profile follows the reference solution quite well for all shown mesh refinements, with a visible discrepancy only within the diffuse interface.
V3 yields an increase in the temperature in the whole domain, which is more pronounced in the domain with a low value for the volume-specific heat capacity.
This increase in the overall temperature distribution is attributed to the increased temperature rate of V3, as shown in the bottom right panel of Fig.~\ref{fig:interface_heat_source_by_capacity_asymdelta}.
The change in temperature rate does not affect the energy rate, as discussed in Appendix~\ref{app:appendix_CSF_energy_rate}.
Essentially, the difference in the interpolation yields a visible decrease of the effective volume-specific heat capacity $\volumetricCapacityEff$ across the interface; see the top right panel of Fig.~\ref{fig:interface_heat_source_by_capacity_asymdelta}.
This results in a higher temperature rate at the same energy rate as V1 and a higher overall temperature because the spatial average of the heat capacity is slightly lower.
For both cases, the temperature profiles are much closer to the reference solution as compared to the classical CSF, shown in the left panel of Fig.~\ref{fig:onedim_symdelta_dyn_stat_h_stud_consteps_1e-6_T-profiles}.
As desired, the modeling error introduced by approximating the sharp interface problem with a CSF approach could be significantly reduced by using the proposed parameter-scaled delta functions compared to the standard delta function.
Moreover, the interpolation of the volume-specific heat capacity $\volumetricCapacityEff(\indicator)$ as one single parameter seems to additionally reduce this modeling error as compared to a separate interpolation of the parameters $\rhoEff(\indicator)$ and $\cpEff(\indicator)$ (with $\volumetricCapacityEff(\indicator) = \rhoEff(\indicator)\,\cpEff(\indicator)$).

Fig.~\ref{fig:onedim_asymdelta_dyn_stat_h_stud_consteps_1e-6_T-error} shows the relative temperature error to the sharp interface reference solution for the different mesh refinements, including the cases shown in Fig.~\ref{fig:onedim_asymdelta_dyn_stat_h_stud_consteps_1e-6_T-profiles}, indicated by the element size $h$, at the constant interface thickness of $\interfaceThickness = \SI{6}{\mu m}$.
The number of finite elements across the interface $\numberOfElementsInInterface$ \eqref{eq:numberOfElementsInInterface} is annotated.
In comparison with the classical CSF, shown in the left panel of Fig.~\ref{fig:onedim_symdelta_dyn_stat_h_stud_consteps_1e-6_T-error}, the relative temperature error follows a similar profile with an asymptotic trend, but the magnitude of the relative temperature error is approximately one order lower for V3 and almost two orders lower for V1.
Doubling the number of finite elements across the interface to $\numberOfElementsInInterface = 64$ changes the relative temperature error by less than \num{1}\% for V1.
At $\numberOfElementsInInterface = 16$, the relative temperature error to the reference solution already meets the tolerance level of \num{1}\% for the chosen interface thickness of $\interfaceThickness = \SI{6}{\mu m}$.
While the difference between two discretizations with the same interface thickness can be attributed to the spatial discretization error, the remaining error in the asymptotic limit of fine discretizations represents the modeling error of the diffuse interface approach.
In sum, these results confirm that the parameter-scaled CSF performs significantly better at the given diffuse interface thickness compared to the classical CSF, which never meets \num{1}\% tolerance at this interface thickness.
For the case V3, $\numberOfElementsInInterface = 64$ finite elements across the interface are required to attain a change in the relative temperature error by less than \num{1}\%.

Fig.~\ref{fig:onedim_asymdelta_dyn_stat_interface_width_stud_ni_128_T-error} shows the $L^{2}$-norm of relative temperature error for different values of the interface thickness $\interfaceThickness$ and for different discretizations of the interface region.
For V1, the relative temperature error decreases with a convergence rate of order $O(\interfaceThickness^{1})$ with respect to the interface thickness for small values of the interface thickness.
To achieve a \num{1}\% tolerance, using V1 with $\numberOfElementsInInterface = 32$, the interface thickness has to obey $\interfaceThickness < \SI{10.3}{\mu m}$ or \num{10.3}\% of the length parameter $a$, which is about \num{41} times less restrictive compared to the result obtained by the classical CSF model, see Section~\ref{sec:classicalCSFexample}.
For V3, the convergence behavior of the relative temperature error with respect to the interface thickness only reaches the order $O(\interfaceThickness^{1})$ for a sufficient discretization of the interface region.
If the interface is insufficiently discretized, the convergence rate tends to decrease.
This may be due to the fact that the distribution of the volume-specific heat capacity $\volumetricCapacityEff$ is not centered around the interface midplane, as can be seen in the top right panel of Fig.~\ref{fig:interface_heat_source_by_capacity_asymdelta}.
For V3 with $\numberOfElementsInInterface = 64$, the tolerance of \num{1}\% is achieved with an interface thickness of $\interfaceThickness < \SI{2.4}{\mu m}$ or \num{2.4}\% of $a$, about ten times less restrictive than the classical CSF model in Section~\ref{sec:classicalCSFexample}.
Using a homogeneous mesh in the whole 1D domain that supports a sufficiently small interface thickness sufficiently resolved, for V1 622 finite elements and for V3 5,338 finite elements are required.
Both cases show a significant improvement compared to the classical CSF, where the corresponding discretization results in 51,625 finite elements.
Note that for V1, the 1\% tolerance is attained in a range of the interface thickness, where the convergence order of $O(\interfaceThickness^1)$ is not yet established.
For reaching a tolerance of \num{0.1}\%, the interface thickness has to be less than $\interfaceThickness < \SI{0.32}{\mu m}$ for V1 and $\interfaceThickness < \SI{0.188}{\mu m}$ for V3.
Here, the convergence order of $O(\interfaceThickness^1)$ is attained for V1.
In comparison, the classical CSF model would require an interface thickness of $\interfaceThickness < \SI{0.0245}{\mu m}$ to reach the \num{0.1}\% tolerance.
This value is extrapolated from the data shown in the left panel of Fig.~\ref{fig:onedim_symdelta_dyn_stat_interface_width_stud_ni_128_T-error}, assuming the linear trend holds.
Here, the improvement is a reduction in interface thickness by a factor of 13 for the V1 case and by a factor of 8 for the V3 case.

In conclusion, the introduced parameter-scaled CSF model performs significantly better than the classical CSF, discussed in Section~\ref{sec:classicalCSFexample}.
For subsequent experiments, we use $\numberOfElementsInInterface = 128$, which aims to minimize the errors due to inadequate resolution across the interface, which reduces the number of free parameters in the investigations.
From Fig.~\ref{fig:onedim_asymdelta_dyn_stat_h_stud_consteps_1e-6_T-error} and Fig.~\ref{fig:onedim_asymdelta_dyn_stat_interface_width_stud_ni_128_T-error}, one could conclude that using less resolution across the interface, e.g., $\numberOfElementsInInterface = 8$ or $\numberOfElementsInInterface = 16$, might be a more efficient balance between mesh resolution and accuracy; in fact, the 3D example in Section~\ref{sec:applcation_meltpool} below uses $\numberOfElementsInInterface \approx 10$.
The criterion for the required interface thickness to predict the temperature field with a given level of accuracy is less restrictive by at least one order of magnitude for the proposed parameter-scaled CSF procedure compared to classical CSF approaches.
This has the potential to drastically reduce computational costs because a larger interface region can accurately be discretized with larger finite elements, which allows for much coarser finite element meshes and the use of larger time step sizes.
Notably, in higher dimensions and with regular element aspect ratios, the savings become even more significant.
For 3D problems with a constant discretization resolution in the interface thickness direction, the number of points within the diffuse interface region scales quadratically with the interface thickness, underlining the importance of this result.
For high heat capacity ratios, the classical CSF amplifies the temperature in the domain with a low heat capacity.
This leads to challenges in the modeling of temperature-dependent interface fluxes, e.g., evaporation-induced effects, which are discussed in the following section.
The parameter-scaled CSF can reduce such temperature changes across the interface, with case V1 providing the best result.
Thus, we will only consider case V1 for modeling interface heat fluxes in the following.

\section{Consistent formulation of temperature-dependent continuum surface fluxes with improved accuracy}
\label{sec:onedimEvaporation}

A multitude of physical effects govern the dynamics of the melt pool during \PBFAM \cite{meier2017thermophysical,meier2021physics}.
The laser heat source drives the rapid temperature rise, especially at the metal surface where the laser impacts.
When reaching the boiling temperature, evaporation effects emerge at the interface and dominate the melt pool dynamics \cite{bitharas2022interplay}.
Evaporation is characterized by a mass flux across the interface, the intensity of which is determined by the interface temperature.
Specifically for heat transfer, the following two effects have a significant impact to be investigated in the following:
First, the vapor mass flow induces evaporation-induced cooling due to the latent heat of the phase change from liquid to gas, modeled as a temperature-dependent interface flux.
Second, the evaporative mass flux causes convective heat transfer.
In a sharp interface setting, the jump in material properties, such as conductivity, typically yields a jump in the heat flux and accordingly a kink in the temperature profile.
This kink results in a temperature peak with steep temperature gradients, given the extreme interface heat sources and high ratios in material properties typical for \PBFAM.
Diffusely approximating that sharp peak usually yields steep gradients within the diffuse interface region.
Hence, the temperature can vary significantly across the finite-thickness interface region in regularized models, and diligent care is required to compute regularized interface fluxes that depend on the temperature with high accuracy.

In the following, we present a novel approach, the \emph{interface value} (IV) method, for computing temperature-dependent regularized interface fluxes based on temperature evaluation at the interface midplane, aiming at improved temperature predictions.
We evaluate its accuracy compared to a standard, \emph{continuous evaluation} (CE) method using local temperature values across the interface thickness.
For the evaluation, we extend the two-phase heat transfer simulations, discussed in Sections~\ref{sec:classicalCSF}-\ref{sec:parameterScaledCSF}, by evaporation-induced effects, representing the thermal behavior of melt pool dynamics.
As an additional measure of accuracy with regard to the fully coupled thermo-hydrodynamic melt pool problem to be presented in Section~\ref{sec:applcation_meltpool}, we compute the evaporation-induced recoil pressure from the temperature in a post-processing step for this analysis.

\subsection{Formulation of a consistent interface midplane temperature-dependent continuum surface flux model}
\label{sec:onedimEvaporationEval}

In a sharp interface setting, the temperature is evaluated directly at the interface to compute the temperature-dependent surface flux, making the calculation trivial.
However, in diffuse interface models with temperature-dependent regularized surface fluxes, where a finite interface thickness is introduced over which the temperature varies, different evaluation possibilities arise.
In this section, we present a novel approach by restricting the temperature input to the interface midplane to improve the accuracy of regularized temperature-dependent surface fluxes.
This approach is inspired by existing curvature evaluation approaches for regularized surface tension computations in two-phase flow models to reduce spurious currents \cite{meland2007reduction,zahedi2012spurious}.

A temperature-dependent interface flux $\interfaceFlux(T)$ is assumed to be modeled using a CSF approach.
In the interface value (IV) method, we propose to compute the temperature-dependent interface flux based on the interface temperature
\begin{align} \label{eq:interace_temperature}
	\Tinterface(\Bx) = T(\Bx_{\Gamma}(\Bx))
	\text{.}
\end{align}
We use closest point projection \cite{coquerelle2016fourth} to determine, for a given point $\Bx$ within the diffuse interface, the associated closest point $\Bx_{\Gamma}(\Bx)$ on the interface midplane $\GammaLG$ at which the interface temperature $T(\Bx_{\Gamma}(\Bx))$ is then evaluated.
This algorithm is described in detail in \cite{schreter2024consistent} in a similar context.
We note that due to the non-local computational procedure required for the projection algorithm, the computational complexity of obtaining the interface temperature is higher compared to the direct local computation.
The algorithmic complexity increases, especially when considering a parallelized MPI-based implementation using domain decomposition and for large meshes, as is the case in 3D.
The employed restriction of the temperature input to the interface midplane, used for computing the temperature-dependent continuum surface flux
\begin{align} \label{eq:interface_value}
	\IvInterfaceFlux(\Bx) = \interfaceFlux(\Tinterface(\Bx))\,\deltaI(\indicator(\Bx))
\end{align}
ensures a constant distribution of the interface temperature across the interface thickness.
The interface value method is denoted with the subscript $(\bullet)_{\*{IV}}$, and $\deltaI(\indicator)$ is the delta function of the chosen CSF modeling approach.

For the evaluation of the IV method, as a more straightforward alternative, we consider a \emph{continuous evaluation} (CE) method, where the local temperature value $T(\Bx)$ within the diffuse interface is used to compute temperature-dependent continuum surface flux distribution:
\begin{align} \label{eq:continuous_evaluation}
	\CeInterfaceFlux(\Bx) = \interfaceFlux(T(\Bx))\,\deltaI(\indicator(\Bx))
	\text{.}
\end{align}
We denote this variant with the subscript $(\bullet)_{\*{CE}}$.

\subsection{Investigation of temperature-dependent CSF modeling for evaporation effects}
\label{sec:onedimEvaporationInvestigation}

In this section, we evaluate the strengths and weaknesses of the IV method and the CE method based on two benchmark cases.
Therefore, we solve the two-phase heat transfer equation according to \eqref{eq:heat_equation} and incorporate evaporation effects relevant for \PBFAM.

Preliminary to the numerical study, we introduce evaporation-related model equations commonly used in the benchmark examples of this section as well as in part in Sections~\ref{sec:twodimFixedMeltPool} and \ref{sec:applcation_meltpool}.
We consider the interface heat flux on the liquid-gas interface $\GammaLG$
\begin{align} \label{eq:laser_plus_evapor_heat_loss}
	\interfaceHeatSource = \qLaserSharp + \qVaporSharp(T)
	\quad \text{ on } \GammaLG
\end{align}
consisting of the laser heat source $\qLaserSharp$, specified individually for the benchmark examples in this section and in Sections~\ref{sec:twodimFixedMeltPool} and \ref{sec:applcation_meltpool}, and the evaporation-induced heat loss $\qVaporSharp(T)$.
According to \cite{meier2021novel}, the evaporation-induced heat loss is defined as
\begin{align} \label{eq:evaporative_heat_loss_wo_specific_enthalpy_sharp}
	\qVaporSharp(T) = -\hv\,\mDot(T)
	\quad \text{ on } \GammaLG
\end{align}
with the specific latent heat of evaporation $\hv$.
We determine the vapor mass flux $\mDot(T)$ at the liquid-gas interface $\GammaLG$ by the model proposed by Knight \cite{knight1979theoretical} and later used by Anisimov and Khokhlov \cite{anisimov1995instabilities} according to
\begin{align} \label{eq:evaporative_mass_flux}
	\mDot(T) = 0.82\,\cs\,\pv(T)\,\sqrt{\frac{M}{2\pi\,R\,T}}
	\quad \text{ on } \GammaLG\,
	\text{,}
\end{align}
with the molar mass $M$ and the molar gas constant $R$.
The sticking constant $\cs$ typically takes a value close to one, i.e., $\cs = 1$ for metals \cite{khairallah2016laser}.
The evaporation-induced recoil pressure $\pv(T)$ is determined via the phenomenological model by Anisimov and Khokhlov \cite{anisimov1995instabilities}:
\begin{align} \label{eq:recoil_pressure}
	\pv(T) = 0.54\,\pa\,\exp\left(-\frac{\hvBar}{R}\left(\frac{1}{T}-\frac{1}{\Tv}\right)\right)
	\quad \text{ on } \GammaLG
	\text{.}
\end{align}
Here, $\pa = \SI{e5}{Pa}$ is the atmospheric pressure, $\hvBar$ is the molar latent heat of evaporation, and $\Tv$ is the boiling temperature.

Instead of consistently resolving the evaporation-induced vapor/gas flow, most existing melt pool models only consider the fluid dynamics within the melt pool and account for the interaction between the melt and vapor phase via phenomenological models for the corresponding thermal and mechanical interface fluxes, applied as Neumann boundary conditions on the melt pool surface \cite{courtois2014complete,fuchs2021sph,khairallah2016laser}.
For the heat transfer equation \eqref{eq:heat_equation}, this means that if we neglect the convective heat transfer resulting from the evaporation-induced vapor/gas flow, i.e., $\Bu\cdot\nabla T=0$, the expression for the evaporation-induced cooling $\qVaporSharp(T)$ \eqref{eq:evaporative_heat_loss_wo_specific_enthalpy_sharp} needs to be adapted for this case.
Since the vapor flow is not resolved in such phenomenological models, an additional term is required to account for the enthalpy transported by the vapor mass flux, defined as:
\begin{align} \label{eq:evaporative_heat_loss_with_specific_enthalpy_sharp}
	\qVaporSharp(T) = -(\hv+ h(T))\,\mDot(T)
	\quad \text{ on } \GammaLG
	\quad \text{with} \quad
	h(T) = \int_{\Thvref}^{T}\cp(\bar{T})\,\diffd\bar{T}
	\text{.}
\end{align}
Here, $\Thvref$ is the reference temperature for the specific enthalpy $h(T)$.

In the benchmark cases discussed within this section, we investigate the heat transfer problem with and without evaporation-induced convective heat transfer, requiring an expression for the evaporation-induced cooling either according to \eqref{eq:evaporative_heat_loss_wo_specific_enthalpy_sharp} or according to \eqref{eq:evaporative_heat_loss_with_specific_enthalpy_sharp}.

The evaporation-induced cooling $\qVaporSharp(T)$ represents a temperature-dependent interface flux.
In a CSF model, these quantities need to be determined within the diffuse interface region $\DiffuseInterfaceRegion$, where the temperature may have varying values in principle.
We distinguish between the two presented variants to evaluate the temperature within the diffuse interface.
The IV method \eqref{eq:interface_value} applied to evaporative cooling reads as
\begin{align}
	\qVaporIV(\Bx) = \qVaporSharp(\Tinterface(\Bx))\,\deltaI(\indicator(\Bx)) \label{eq:evaporative_heat_loss_interface_value}
	\text{.}
\end{align}
The CE method \eqref{eq:continuous_evaluation} applied to evaporative cooling reads as
\begin{align}
	\qVaporCE(\Bx) &= \qVaporSharp(T(\Bx))\,\deltaI(\indicator(\Bx)) \label{eq:evaporative_heat_loss_continuous_evaluation}
	\text{.}
\end{align}
In \eqref{eq:evaporative_heat_loss_interface_value} and \eqref{eq:evaporative_heat_loss_continuous_evaluation}, $\deltaI$ is the respective delta function for the volume-specific heat capacity interpolation type as listed in Table~\ref{tab:parameter_scaled_CSF_cases}.
This choice is made for consistency reasons to have the same delta function for distributing both the laser heat source and the evaporation-induced cooling across the interface thickness.

In a coupled thermo-hydrodynamic model of \PBFAM (cf.~Section~\ref{sec:applcation_meltpool}), the recoil pressure is the dominating mechanical force acting on the melt pool surface.
According to \eqref{eq:recoil_pressure}, it scales exponentially with the temperature, i.e., it is very sensitive with respect to modeling and discretization errors in the temperature field.
Therefore, in the benchmark examples of this section, we calculate the recoil pressure error in a post-processing step.
The recoil pressure force $\fRecoilPressureSharp(T)$ acting on the liquid-gas interface $\GammaLG$ is calculated according to
\begin{align} \label{eq:recoil_pressure_force_sharp}
	\fRecoilPressureSharp(T) = \pv(T) \, \normalLG
	\quad \text{ on } \GammaLG
	\text{,}
\end{align}
where the normal vector $\normalLG$ is the unit normal vector of the interface midplane $\GammaLG$, pointing into $\OmegaL$.
For the continuum surface flux representation of the recoil pressure force, we choose to scale the parameter-scaled delta function with the density because interface forces apply to the momentum equation \eqref{eq:momentum_equation} and the linear momentum is proportional to the density.
This approach is equivalent to the well-established density-scaled delta function $\deltaDensity(\indicator)$, which balances the linear momentum of the incompressible Navier--Stokes equations \cite{yokoi2014density}.
Application of the IV method \eqref{eq:interface_value} to the recoil pressure yields
\begin{align}
	\fRecoilPressureIV(\Bx) = \fRecoilPressureSharp(\Tinterface(\Bx))\,\deltaDensity(\indicator(\Bx)) \label{eq:recoil_pressure_interface_value}
	\text{,}
\end{align}
and similarly of the CE method yields \eqref{eq:continuous_evaluation} in
\begin{align}
	\fRecoilPressureCE(\Bx) &= \underbrace{\pv(T(\Bx))\,\deltaDensity(\indicator(\Bx))}_{\pvDiffuse(\Bx)}\,\normalLG(\Bx) \label{eq:recoil_pressure_continuous_evaluation}
	\text{.}
\end{align}

\subsubsection{Laser-induced heating benchmark example with evaporation-induced cooling}
\label{sec:onedimEvaporationCSFexample}

The 1D benchmark example illustrated in Fig.~\ref{fig:onedim_sketch} and described in Section~\ref{sec:classicalCSFexample} is used to evaluate the influence of evaporation-induced cooling $\qVaporSharp(T)$ \eqref{eq:evaporative_heat_loss_with_specific_enthalpy_sharp} on the temperature.
From the temperature, the phenomenological recoil pressure $\pv(T)$ \eqref{eq:recoil_pressure} is computed in a post-processing step.
In the heat equation \eqref{eq:onedim_heat_equation}, the flux term is determined with the parameter-scaled CSF model $\volumetricHeatSource = \interfaceHeatSource\,\deltaSingleWeighted$ with a delta function weighted proportional to the effective volume-specific heat capacity $\volumetricCapacityEff(\indicator)$ that is interpolated across the interface as one material property using the arithmetic mean interpolation \eqref{eq:parameter_transition_arithmetic}.
Thus, the parameter-scaled CSF corresponds to case V1, as listed in Table~\ref{tab:parameter_scaled_CSF_cases}.
According to \eqref{eq:laser_plus_evapor_heat_loss}, the flux term contains the constant laser heat source of $\qLaserSharp = \SI{e10}{W\per m^2}$ and the evaporation-induced cooling.
We assume that the interface is stationary and the evaporation-induced cooling has to be modeled according to \eqref{eq:evaporative_heat_loss_with_specific_enthalpy_sharp}.
We discuss both the continuous evaluation (CE) \eqref{eq:evaporative_heat_loss_continuous_evaluation} and the interface value (IV) method \eqref{eq:evaporative_heat_loss_interface_value} for modeling the evaporative cooling.
All material parameters are listed in Table~\ref{tab:param_ti64}.
To exclude discretization errors in this investigation, $\numberOfElementsInInterface = 128$ finite elements across the interface thickness are considered.

\begin{figure}[tbp!]
	\centering
	\includegraphics{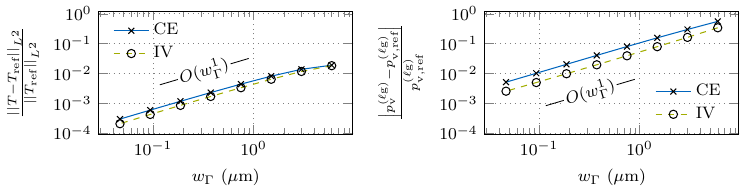}
	\caption{
		Error measures for the instationary temperature profile at $t=\SI{e-5}{s}$ resulting from interface heating with an interface laser heat source of $\interfaceHeatSource = \SI{e10}{W\per m^2}$ and evaporation-induced heat loss according to \eqref{eq:evaporative_heat_loss_with_specific_enthalpy_sharp} using the parameter-scaled CSF approach V1, described in Table~\ref{tab:parameter_scaled_CSF_cases} for the continuous evaluation (CE) method and the interface value (IV) method.
		The interface region is resolved by a constant number of $\numberOfElementsInInterface = 128$ finite elements, ensuring a converged solution with respect to spatial discretization.
		The reference temperature profile $\Tref$ is determined using a sharp interface approach:
		(left) relative error in the temperature profile for different diffuse interface thicknesses $\interfaceThickness$;
		(right) relative error in the phenomenological evaporation-induced recoil pressure \eqref{eq:recoil_pressure} for CE according to \eqref{eq:recoil_pressure_continupus_evaluation_integrated} and for the IV $\pvIV = \pv(\Tinterface)$.
		The reference recoil pressure is determined with the interface value of the reference temperature profile $\pvRef = \pv(\TinterfaceRef)$.
	}
	\label{fig:onedim_asymdelta_rp_interface_width_stud}
\end{figure}

Fig.~\ref{fig:onedim_asymdelta_rp_interface_width_stud} shows the instationary result at $t = \SI{e-5}{s}$ for the CE and IV method.
The relative temperature error with respect to the sharp interface reference solution, shown in the left panel, decreases with a convergence rate of order $O(\interfaceThickness^1)$ for small values of the interface thickness $\interfaceThickness$, which is a similar convergence behavior as the benchmark example without evaporation-induced cooling, discussed in Section~\ref{sec:parameterScaledCSFexample}.
For CE, the interface thickness is required to be less than $\interfaceThickness < \SI{2.01}{\mu m}$ or \num{2.01}\% of the length parameter $a$ to achieve a tolerance of \num{1}\%, about five times finer than for the example without evaporation-induced cooling.
A slight improvement is achieved by the IV method, where the interface thickness is required to be less than $\interfaceThickness < \SI{2.54}{\mu m}$ or \num{2.54}\% of $a$, about four times finer than for the example without evaporation effects.
These values, however, are attainted in a range of interface thicknesses, where the convergence order of $O(\interfaceThickness^1)$ is not yet established.
To achieve a tolerance of \num{0.1}\%, for CE, the interface thickness only needs to be twice as fine as the benchmark example without evaporation-induced cooling and about \num{1.5} times finer for the IV method.

The right panel of Fig.~\ref{fig:onedim_asymdelta_rp_interface_width_stud} shows the relative error in the recoil pressure with respect to the reference solution.
To obtain the absolute value of the recoil pressure using CE, we integrate the continuously evaluated recoil pressure $\pv(T)$ in \eqref{eq:recoil_pressure} over the thickness of the diffuse interface:
\begin{align} \label{eq:recoil_pressure_continupus_evaluation_integrated}
	\pvCE = \int_{-\frac{\interfaceThickness}{2}}^{\frac{\interfaceThickness}{2}} \, \pvDiffuse(\distance) \, \diffd\distance
	\text{.}
\end{align}
For IV, we determine the recoil pressure according to \eqref{eq:recoil_pressure} based on the interface value of the temperature $\pvIV = \pv(\Tinterface)$.
The reference solution of the recoil pressure $\pvRef = \pv(\TinterfaceRef)$ is determined based on the interface value of the sharp interface reference temperature solution $\TinterfaceRef = \Tref(x_{\Gamma})$.
According to the plot in the right panel of Fig.~\ref{fig:onedim_asymdelta_rp_interface_width_stud}, the relative error in the recoil pressure decreases with a convergence rate of order $O(\interfaceThickness^{1})$ with respect to the interface thickness for both the CE and the IV method.
CE achieves a tolerance of \num{1}\% with an interface thickness less than $\interfaceThickness < \SI{0.0926}{\mu m}$ or \num{0.1}\% of the length parameter $a$.
This value is \num{22} times smaller than achieving the same accuracy of the relative temperature error, representing a significant decrease in the required interface thickness when changing the error measure.
The \num{1}\% tolerance for the relative error in the recoil pressure is achieved at an interface thickness of $\interfaceThickness < \SI{0.192}{\mu m}$ or \num{0.2}\% of $a$ when using the IV method.
Here, the decrease in the required interface thickness due to the change in the error measure is approx.~\num{13} times.

In conclusion, introducing evaporation effects demands a much finer spatial discretization of the interface region since the exponential nature of the formulation for the phenomenological recoil pressure \eqref{eq:recoil_pressure} requires a precise temperature at the interface.
Since the thermal problem is driven by an interface heat flux modeled by CSF, the temperature profile in the diffuse interface region is subject to a significant modeling error due to the diffuse interface assumption and an additional spatial discretization error.
It is demonstrated that accuracy gains can be achieved for predicted temperature-dependent interface fluxes if the temperature is evaluated at the interface midplane instead of using local values across the interface thickness.

\subsubsection{Laser-induced heating benchmark example with evaporation-induced cooling and convective heat transfer}
\label{sec:onedimConvectionCSFexample}

In a coupled thermo-hydrodynamic melt pool simulation, taking into account evaporation-induced flow, convective heat transfer occurs due to the evaporation-induced velocity field in the gas domain.
The velocity exhibits a jump at the interface due to the phase transition, where the fluid density decreases by orders of magnitude across the interface as the metal evaporates.

In this example, we study the effect of the evaporation-induced convective heat transfer based on the benchmark example illustrated in Fig.~\ref{fig:onedim_sketch}.
Therefore, we consider the heat equation~\eqref{eq:heat_equation} in the 1D form:
\begin{align} \label{eq:onedim_heat_equation_with_convection}
	\underbrace{\interp{\left(\rho\cp\right)}}_{\volumetricCapacityEff} \left( \fracPartial{T}{t} + u\,\fracPartial{T}{x} \right) = \fracPartial{}{x} \left(\interp{\conductivity}\,\fracPartial{T}{x}\right) + \volumetricHeatSource
	\quad\text{ in }\Omega\times[0,t]
	\text{.}
\end{align}
Based on the simplifying assumption of an incompressible flow, the requirement of mass conservation directly yields the result that the velocity is inversely proportional to the effective density \cite{schreter2024consistent}.
In our example, the evaporated volume is assumed to be compensated by a prescribed inflow velocity on the liquid side of the interface to yield a spatially fixed interface location.
With these assumptions, the evaporation-induced convection velocity in the 1D domain can be analytically calculated as
\begin{align} \label{eq:onedim_evapor_velocity_profile}
	u(x) = \frac{\mDot(\Tinterface)}{\harmInterp{\rho}\left(\indicator(x)\right)}
	\quad \text{ in } \Omega
	\text{.}
\end{align}
For calculating the velocity, we apply a density interpolation based on the harmonic mean $\harmInterp{\rho}$ \eqref{eq:parameter_transition_harmonic}, as suggested in \cite{lee2017direct,schreter2024consistent} to satisfy the conservation of mass in diffuse interface models for incompressible two-phase flow with resolved evaporation.
The convection velocity \eqref{eq:onedim_evapor_velocity_profile} is calculated from the temperature at the interface midplane $\Tinterface = T(x_{\Gamma})$.
The heat flux $\volumetricHeatSource$ in the heat equation \eqref{eq:onedim_heat_equation_with_convection} contains the laser heat source ${\qLaserSharp = \SI{e10}{W\per m^2}}$ and the evaporation-induced heat loss $\qVaporIV$ \eqref{eq:evaporative_heat_loss_interface_value} for which we use the IV method as the most accurate method from Section~\ref{sec:onedimEvaporationCSFexample}.
Since we consider evaporation-induced convective heat transfer across the interface by \eqref{eq:onedim_evapor_velocity_profile}, the evaporation-induced cooling is determined by \eqref{eq:evaporative_heat_loss_wo_specific_enthalpy_sharp}, without considering the specific enthalpy term, in contrast to the example in Section~\ref{sec:onedimEvaporationCSFexample}.
The heat fluxes are modeled as volume fluxes in the interface region via the parameter-scaled CSF method $\volumetricHeatSource = \interfaceHeatSource\,\deltaSingleWeighted$, weighted proportional to the effective volume-specific heat capacity $\volumetricCapacityEff$.
The volume-specific heat capacity is interpolated across the interface as one material property using the arithmetic mean interpolation \eqref{eq:parameter_transition_arithmetic} $\volumetricCapacityEff = \volumetricCapacityArith$
Thus, the parameter-scaled CSF formulation is equivalent to the case V1, described in Table~\ref{tab:parameter_scaled_CSF_cases}.
The remaining problem description is adopted from Section~\ref{sec:classicalCSFexample}, and the material parameters are listed in Table~\ref{tab:param_ti64}.

\begin{figure}[bp!]
	\centering
	\includegraphics{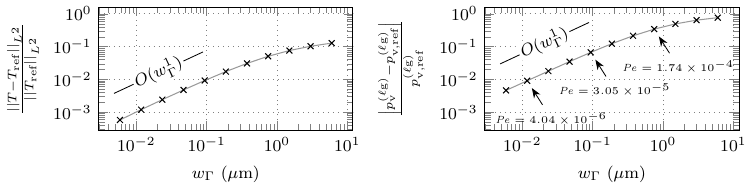}
	\caption{
		Error measures for the instationary temperature profile at $t=\SI{e-5}{s}$ resulting from interface heating with an interface laser heat source of $\interfaceHeatSource = \SI{e10}{W\per m^2}$ and evaporation-induced heat loss according to \eqref{eq:evaporative_heat_loss_interface_value} using the parameter-scaled CSF approach V1, described in Table~\ref{tab:parameter_scaled_CSF_cases} and the interface value (IV) method.
		The interface region is resolved by a constant number of $\numberOfElementsInInterface = 128$ finite elements, ensuring a converged solution with respect to spatial discretization.
		The reference temperature profile $\Tref$ is determined using a sharp interface approach:
		(left) relative error in the temperature profile for different diffuse interface thicknesses $\interfaceThickness$;
		(right) relative error in the phenomenological evaporation-induced recoil pressure \eqref{eq:recoil_pressure} using the IV mothod $\pvIV = \pv(\Tinterface)$.
		The element Péclet number $\Pe$ \eqref{eq:peclet_number} of the gas phase is annotated.
	}
	\label{fig:onedim_asymdelta_evapor_interface_width_stud}
\end{figure}

In Fig.~\ref{fig:onedim_asymdelta_evapor_interface_width_stud}, the instationary result at $t = \SI{e-5}{s}$ is shown with relative error measures compared to the reference solution.
The reference solution is determined with a sharp interface model, and the velocity profile $u(x)$ used to determine the reference solution is constant within the phases, namely $u(x) = \mDot(\TinterfaceRef) / \inGas{\rho}$ in the gas domain $\OmegaG$ and $u(x) = \mDot(\TinterfaceRef) / \inLiquid{\rho}$ in the liquid domain $\OmegaL$ with a jump at the interface.
The relative temperature error to the sharp interface reference solution is shown in the left panel.
With a convergence order of $O(\interfaceThickness^{1})$, the error decreases for small values of the interface thickness.
The interface thickness has to be smaller than ${\interfaceThickness < \SI{0.103}{\mu m}}$ or \num{0.1}\% of the length parameter $a$ to achieve a tolerance of \num{1}\%, which is smaller by a factor of \num{25} as compared to the case without the convection in Section~\ref{sec:onedimEvaporationCSFexample}.

The right panel in Fig.~\ref{fig:onedim_asymdelta_evapor_interface_width_stud} shows the relative error in the recoil pressure with respect to the sharp interface reference solution.
Here, the convergence order of $O(\interfaceThickness^{1})$ is attained for small values of the interface thickness, but the tolerance of \num{1}\% is only reached with an interface thickness of $\interfaceThickness < \SI{0.0129}{\mu m}$ or \num{0.013}\% of $a$, which is \num{15} times finer, than reaching the same tolerance in the example without convective heat transfer in Section~\ref{sec:onedimEvaporationCSFexample}.

Introducing the convective heat transfer across the interface, resulting from the vapor mass flux, makes the model more complex, and the interface thickness needs to be refined to remain at the same level of accuracy.
Temperature-dependent interface fluxes, such as the recoil pressure, require a precise interface temperature.
In the diffuse interface region, the error introduced by the CSF model persists, and a significant refinement of the interface thickness is required to retain the tolerance in the interface temperature.

The element Péclet number $\Pe$ describes the ratio between convective heat transfer and conductive heat transfer in an element and is defined as follows:
\begin{align} \label{eq:peclet_number}
	\Pe = \frac{\rho\cp\,u\,h}{\conductivity}
	\text{.}
\end{align}
In the plot in the right panel of Fig.~\ref{fig:onedim_asymdelta_evapor_interface_width_stud}, the element Péclet numbers $\Pe$ of the gas phase are given for some simulations with different interface thicknesses.
The Péclet number is very small, indicating that the problem is dominated by conduction in element scale.
Since the problem is not dominated by convection, no stabilization is required.

\section{Benchmark example: laser-induced heating of a 2D fixed melt pool surface}
\label{sec:twodimFixedMeltPool}

\begin{figure}[bp!]
	\centering
	\includegraphics{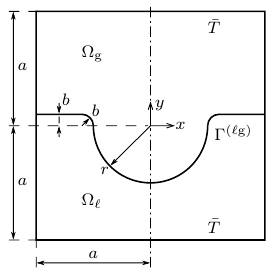}
	\includegraphics[width=0.305\textwidth,trim={1.35cm -2.9cm 1.35cm 1.35cm},clip]{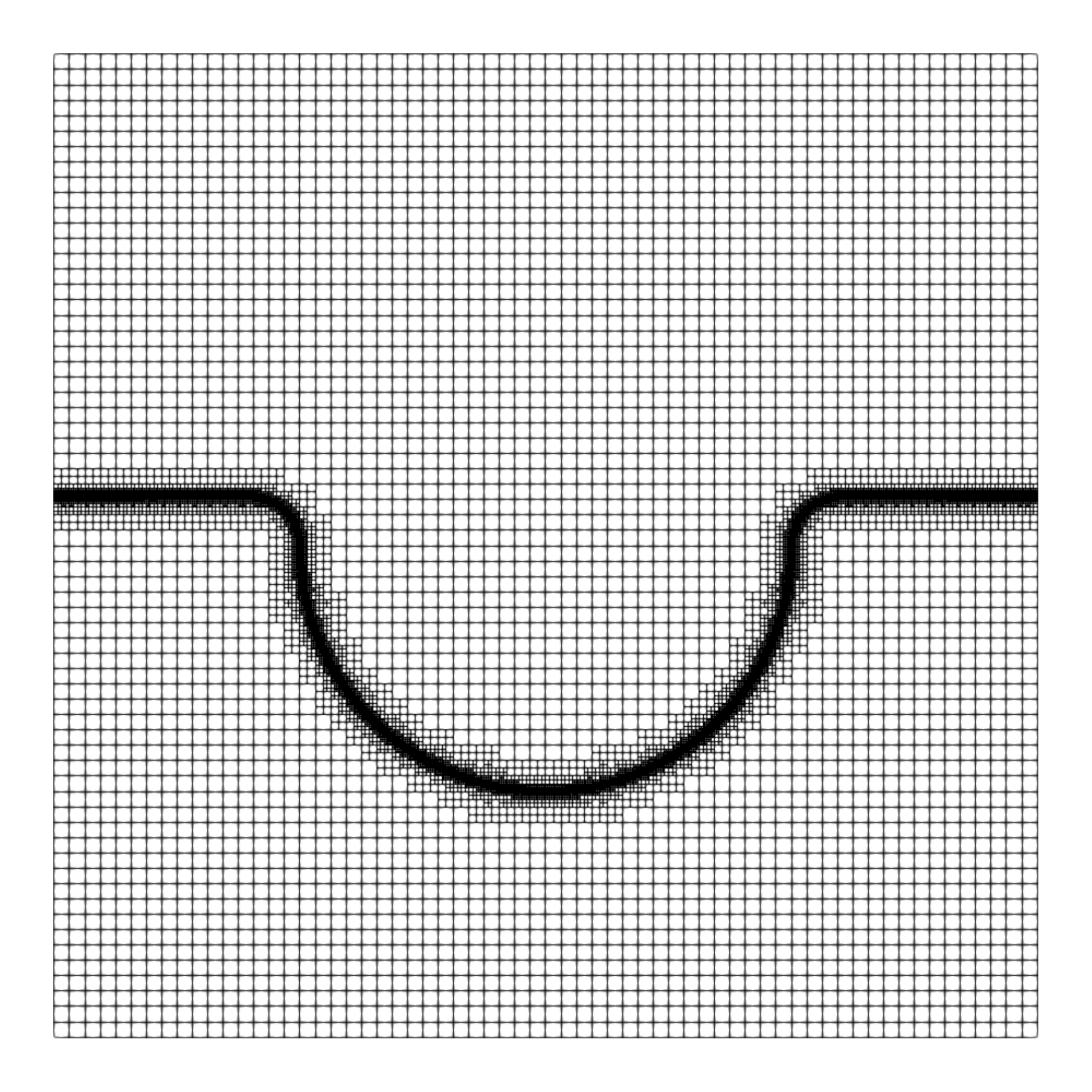}
	\hspace*{1mm}
	\includegraphics[width=0.305\textwidth,trim={1.35cm -2.9cm 1.35cm 1.35cm},clip]{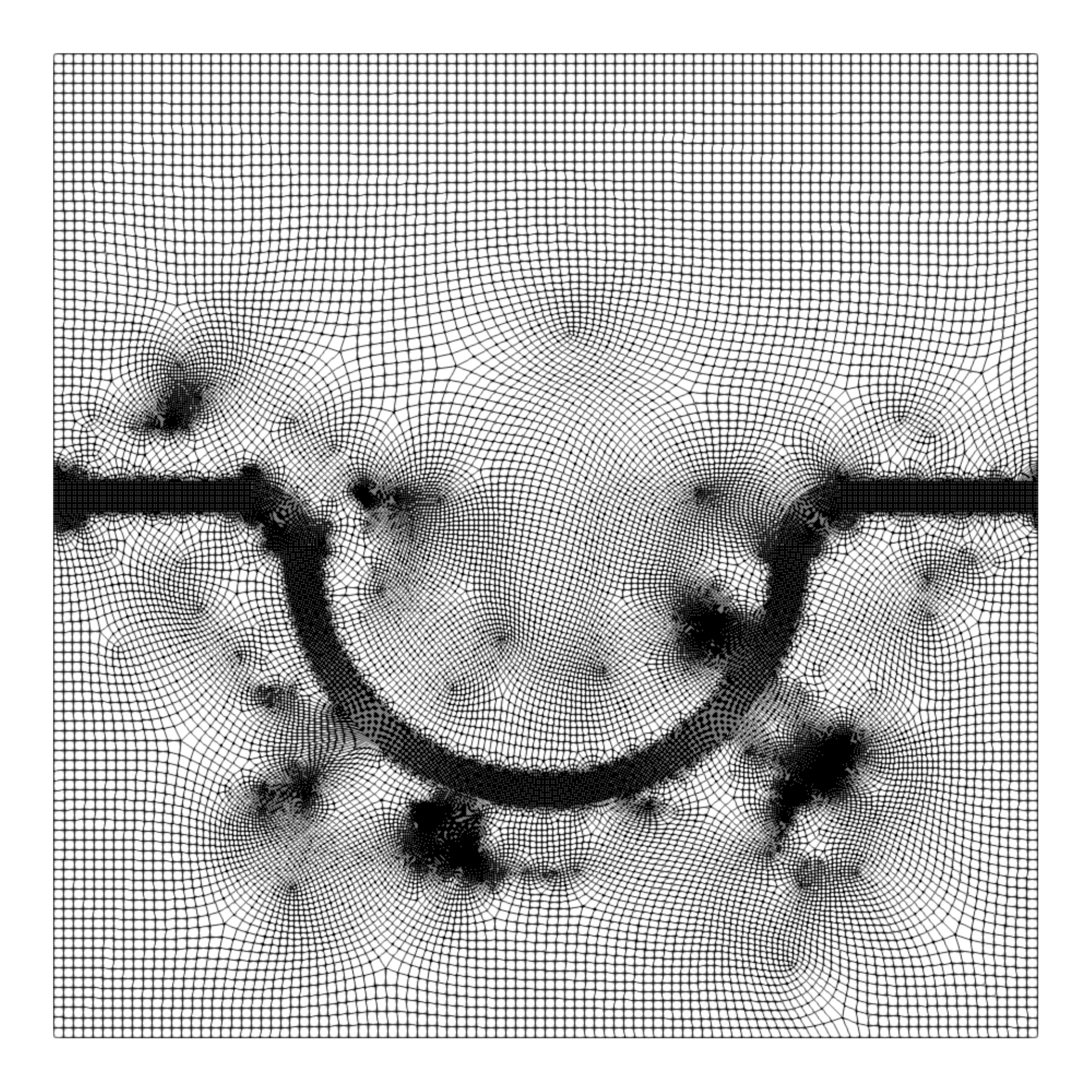}
	\caption{
		2D fixed melt pool surface:
		(left) sketch;
		(center) Cartesian finite element mesh of the small interface approach with local refinement in the interface region.
		The coarser mesh of the approach with the large interface thickness is indistinguishable at the shown scale due to local mesh refinement.
		(right) finite element mesh with element edges coinciding with the interface midplane $\GammaLG$.
		}
	\label{fig:fixed_melt_pool_sketch}
\end{figure}

In the following benchmark example, the parameter-scaled CSF and IV methods are applied to a 2D domain with an interface geometry that mimics a \PBFAM melt pool surface.
Since this study focuses on modeling the laser heat source term in the case of curved interfaces, a spatially fixed interface geometry and a vanishing velocity field in the entire problem domain are considered.
A concave, semi-circular interface with rounded edges represents a vapor depression.
The left panel of Fig.~\ref{fig:fixed_melt_pool_sketch} shows a schematic sketch of the setup.
The domain $\Omega = \{\Bx \in [-a,a]^2\}$ with a length parameter $a = \SI{100}{\mu m}$ is occupied by a gas phase $\OmegaG$ and a liquid phase $\OmegaL$ that are separated by the interface.
The interface midplane $\GammaLG$ is symmetric about the $y$-axis and characterized by a center radius $r = \SI{50}{\mu m}$ and a bead radius $b = \SI{10}{\mu m}$ as shown in the left panel of Fig.~\ref{fig:fixed_melt_pool_sketch}.
We consider the conductive heat transfer according to the heat equation \eqref{eq:heat_equation}.
No convective heat transfer is considered as the velocity $\Bu$ remains zero.
The indicator $\indicator$ is defined by \eqref{eq:indicator} using the signed distance $\distance(\Bx)$ to the interface midplane $\GammaLG$, which is negative in $\OmegaG$ and positive in $\OmegaL$ according to:
\begin{align} \label{eq:fixed_signed_distance}
	\distance(\Bx) =
	\begin{cases}
		||\Bx||-r
			&\text{for} \quad |x| < r+b \; \wedge \; y < 0 \\
		\min\left\{||\Bx||-r, b-y\right\}
			&\text{for} \quad |x| \geq r+b \; \wedge \; y < 0 \\
		b-y
			&\text{for} \quad |x| \geq r+b \; \wedge \; y \geq 0 \\
		b-\sqrt{(r+b - |x|)^2 + y^2}
			&\text{for} \quad |x| < r+b \; \wedge \; y \geq 0 \\
	\end{cases}
	\text{.}
\end{align}
The diffuse interface region $\DiffuseInterfaceRegion$ is characterized as a narrow band centered around the interface midplane $\GammaLG$ with a thickness corresponding to the interface thickness $\interfaceThickness$.
Typical material parameter values for \tiSixFour are employed and are listed in Table~\ref{tab:param_ti64}.
The volume-specific heat capacity $\volumetricCapacityEff(\indicator)$ is interpolated across the interface as one material property corresponding to case V1 listed in Table~\ref{tab:parameter_scaled_CSF_cases}.
The liquid-gas interface is subject to an interface heat flux $\interfaceHeatSource$ that, according to \eqref{eq:laser_plus_evapor_heat_loss}, comprises the laser heat source $\qLaserSharp(\Bx)$ and the evaporation-induced cooling $\qVaporSharp(T)$.
No vapor flow is resolved in this example, i.e., the evaporation-induced cooling is modeled according to \eqref{eq:evaporative_heat_loss_with_specific_enthalpy_sharp}.
Using the parameter-scaled CSF, the interface heat source $\interfaceHeatSource$ is modeled as a volumetric heat flux $\volumetricHeatSource = \interfaceHeatSource\,\deltaSingleWeighted(\indicator)$ within the diffuse interface region $\DiffuseInterfaceRegion$.
The laser heat source models a spatially fixed laser with a Gaussian profile \cite{meier2021novel}
\begin{align}
	\label{eq:laser_gauss_sharp}
	\qLaserSharp(\Bx) = \absorptivity\,\laserPower\,\frac{2}{\laserRadius^2 \pi}\,\langle\normalLG\laserDirection\rangle \exp\left(-2\left(\frac{\ofLaser{d}(\Bx)}{\laserRadius}\right)^2\right) \quad \text{on } \GammaLG
\end{align}
with the absorptivity $\absorptivity$, the laser power $\laserPower = \SI{250}{W}$, and the laser radius $\laserRadius = \SI{70}{\mu m}$.
$\ofLaser{d}(\Bx)$ is the distance between the point $\Bx$ and the laser beam center line defined by the laser position $\laserPosition$ at the origin and the laser direction $\laserDirection$ corresponding to the negative $y$-direction.
The Macauley bracket $\langle\bullet\rangle$ yields the argument's value for positive inputs and zero otherwise.
Initially, the temperature is uniform at $T_{0} = \SI{500}{K}$.
At the top and bottom boundaries, the temperature is prescribed to $\bar{T} = \SI{500}{K}$ \eqref{eq:heat_equation_dirichlet}, and the left and right boundaries are adiabatic according to \eqref{eq:heat_equation_neumann}.

The temperature field is computed by solving the heat equation using the FEM with bilinear quadrilateral elements and the implicit Euler time integration scheme.
We employ a Cartesian grid with a base element size of $h_{\*{max}} = \SI{3.125}{\mu m}$ and use local mesh refinement in the vicinity of the diffuse interface region.

Four different approaches are discussed in the following.
In the first approach, we employ the CE method to compute the evaporation-induced cooling $\qVaporSharp(T)$ according to \eqref{eq:evaporative_heat_loss_continuous_evaluation} along with a small interface thickness of $\interfaceThickness = \SI{0.1}{\mu m}$ and consistently a fine mesh for a sufficient interface resolution.
Here, the finite element size near the interface is locally refined to $h_{\*{min}} = \SI{3.052e-3}{\mu m}$ to have approximately $\numberOfElementsInInterface \approx$ 32 finite elements across the interface, ensuring a converged solution with respect to spatial discretization.
The resulting mesh containing 4,744,480 cells is shown in the center panel of Fig.~\ref{fig:fixed_melt_pool_sketch}.
The 1D benchmark example in Section~\ref{sec:onedimEvaporationCSFexample}, which employed the same governing equations and the same parameter-scaled CSF case, informed the choice of the spatial discretization parameters to obtain an accuracy of 1\% for the recoil pressure.

For the second approach, we employ the IV method to compute the evaporation-induced cooling $\qVaporSharp(T)$ according to \eqref{eq:evaporative_heat_loss_interface_value} and an adequate interface thickness of $\interfaceThickness = \SI{0.2}{\mu m}$, according to the 1D benchmark example, see Fig.~\ref{fig:onedim_asymdelta_rp_interface_width_stud}.
To obtain a sufficient spatial resolution of the interface region of approximately $\numberOfElementsInInterface \approx 32$, the finite element size near the interface is locally refined to $h_{\*{min}} = \SI{6.104e-3}{\mu m}$.
The resulting mesh contains 2,367,454 cells.

For the third and fourth approaches, we employ a larger interface thickness, resulting in a coarser mesh where the element size is locally refined to $h_{\*{min}} = \SI{0.3906}{\mu m}$ in the vicinity of the interface.
The element size is chosen to a value that is deemed feasible to be employed in 3D melt pool simulations with reasonable computational effort.
The interface thickness is set to $\interfaceThickness = \SI{12.5}{\mu m}$ resulting in $\numberOfElementsInInterface \approx 32$ finite elements across the interface and 38,619 finite elements in the mesh.
For the third approach, we employ the CE method for computing evaporation-induced cooling $\qVaporSharp(T)$ according to \eqref{eq:evaporative_heat_loss_continuous_evaluation},
and for the fourth approach, we instead employ the IV method according to \eqref{eq:evaporative_heat_loss_interface_value}.
For all approaches, the time step size is set to $\Delta t = \SI{e-9}{s}$, for which the temporal discretization error was checked to be sufficiently small in a convergence study.

The reference solution is determined from a sharp interface method.
A fitted finite element mesh is used where element edges coincide with the interface midplane $\GammaLG$, as shown in the right panel of Fig.~\ref{fig:fixed_melt_pool_sketch}.
Since the interface is fixed, the element edges remain aligned with the interface midplane throughout the simulation for a constant mesh.
The sharp interface flux $\interfaceHeatSource = \qLaserSharp(\Bx) + \qVaporSharp(T)$ from \eqref{eq:laser_gauss_sharp} and \eqref{eq:evaporative_heat_loss_with_specific_enthalpy_sharp} is applied to the element egdes coinciding with $\GammaLG$.
In the vicinity of the interface, the element size is approximately $h \approx \SI{0.047}{\mu m}$; in the far field, it is approximately $h \approx \SI{2}{\mu m}$.
A time step size of $\Delta t = \SI{e-10}{s}$ is employed.
In a convergence study, the spatial and temporal discretization errors were checked to be sufficiently small.

\begin{figure}[tbp!]
	\centering
	\includegraphics{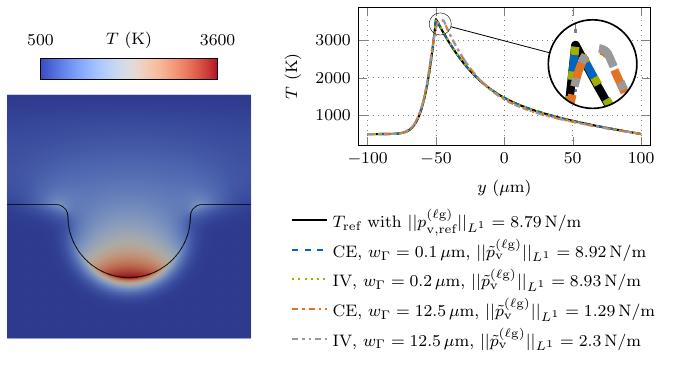}
	\caption{
		Temperature field solution of the 2D fixed melt pool surface benchmark example at $t = \SI{e-5}{s}$:
		(left) temperature field solution obtained for the small interface thickness using the CE method for the evaporation-induced cooling \eqref{eq:evaporative_heat_loss_continuous_evaluation} in the domain $\Omega$ with the interface midplane $\GammaLG$ indicated with a black contour;
		(right) temperature profiles for all approaches along the $y$-axis, i.e., the vertical center line of the domain, compared to the sharp interface reference solution.
		}
	\label{fig:fixed_temperature_distribution}
\end{figure}

The instationary temperature solution at $t = \SI{e-5}{s}$ is shown in Fig.~\ref{fig:fixed_temperature_distribution}.
For all approaches, the peak temperature remains close to the metal surface rather than artificially heating the ambient gas, even in the concave cavity of the melt pool vapor depression.
In the following, we compute the relative temperature error based on the $L^{2}$-norm \eqref{eq:L2norm}, i.e., $||T - \Tref||_{L^{2}} / ||\Tref||_{L^{2}}$.
To assess the recoil pressure error, we compare the $L^{1}$-norm of the recoil pressure distribution between the diffuse interface approaches and the sharp reference solution; the latter is computed to \mbox{$||\pvRef||_{L^{1}} = \int_{\GammaLG} \pv(\Tref(\Bx))\diffd\Bx = \SI{8.79}{N\per m}$} with \eqref{eq:recoil_pressure}.
For the diffuse interface approaches we compute the $L^{1}$-norm of the recoil pressure according to \mbox{$||\pvDiffuse||_{L^{1}} = \int_{\Omega} \pvDiffuse \diffd\Bx$} with \eqref{eq:recoil_pressure_continuous_evaluation}.
The temperature profile using the small interface thickness replicates the reference solution very accurately with a relative error of \num{0.358}\% for the first approach with the CE method and \num{0.359}\% for the second approach with the IV method.
Here, the $L^{1}$-norm of the recoil pressure is \mbox{$||\pvDiffuse||_{L^{1}} = \SI{8.92}{N\per m}$} and \mbox{$||\pvDiffuse||_{L^{1}} = \SI{8.93}{N\per m}$} each, which deviates by \num{1.4}\% and \num{1.6}\% from the reference solution, respectively.
Both error measures are in the same order of magnitude as those obtained for the 1D benchmark example in Section~\ref{sec:onedimEvaporationCSFexample} with similar discretization parameters.
The accuracy is slightly degraded due to the increased complexity of the 2D curved interface geometry.
For the approaches with the large interface thickness, the relative temperature error to the reference solution is \num{3.71}\% for the approach with the CE method and \num{2.05}\% for the approach with the IV method.
Considering that the number of finite elements was cut by a considerable factor compared to the small interface thickness approaches, the accuracy of the temperature is still relatively high.
However, the approach with the large interface thickness and the CE method underestimate the recoil pressure by \num{85.3}\% to $||\pvDiffuse||_{L^{1}} = \SI{1.29}{N\per m}$.
Using the IV method along with the large interface yields a recoil pressure that is underestimated by \num{73.8}\% to $||\pvDiffuse||_{L^{1}} = \SI{2.3}{N\per m}$.
Due to the relatively large interface thickness, the interface temperature is inaccurate, and the exponential nature of the formulation for the phenomenological recoil pressure \eqref{eq:recoil_pressure} amplifies the inaccuracy.
Although the temperature profile appears to be accurate for most of the domain, the large error in the evaporation-induced recoil pressure confirms that the accuracy is mainly influenced by the temperature in the interface region.

The benchmark example in this section increases the complexity of the example in Section~\ref{sec:onedimEvaporationCSFexample} by raising the dimensionality to two and having a curved geometry that mimics the shape of the melt pool while employing the same governing equations and diffuse interface methods.
It is found that the parameter-scaled CSF translates well to higher dimensions and helps to obtain a highly accurate temperature field while employing an adequate interface thickness.
Besides investigating the interface thickness that is sufficient according to the 1D studies in Section~\ref{sec:onedimEvaporationCSFexample}, we investigated approaches with a spatial resolution that we deem feasible for 3D melt pool thermo-hydrodynamics simulations with reasonable computational effort.
With the resulting large interface thickness, the error in the temperature profile seems relatively small when considering the entire domain.
However, even though the temperature accuracy is still relatively high, it leads to a large discrepancy in the recoil pressure due to the exponential relationship between temperature and recoil pressure.

\section{Application of the parameter-scaled CSF model to a melt pool thermo-hydrodynamics simulation}
\label{sec:applcation_meltpool}

In the following, the parameter-scaled CSF method is employed in a 3D, fully coupled melt pool thermo-hydrodynamics simulation.
Based on this proof-of-principle experiment, we show that the parameter-scaled CSF method provides a robust framework for 3D simulations of melt pool dynamics using realistic parameter values and temperature regimes.
For example, using the classical CSF models for interface flux modeling, we could not achieve convergence of the nonlinear Navier-Stokes and/or the nonlinear heat transfer solver, considering the high material property ratios between the phases.
Stationary laser-induced heating of a bare \tiSixFour plate is considered, recreating the experimental setup by Cunningham et al.~\cite{cunningham2019keyhole}.
A similar scenario was investigated in \cite{meier2021novel} with a smoothed particle hydrodynamics framework.

For the heat transfer in the melt pool, we consider the heat equation \eqref{eq:heat_equation}.
The volume-specific heat capacity $\volumetricCapacityEff$ is interpolated across the liquid-gas interface as one material property according to \eqref{eq:parameter_transition_arithmetic}, corresponding to case V1 listed in Table~\ref{tab:parameter_scaled_CSF_cases}.
The conductivity $\conductivityEff$ is interpolated across the interface according to \eqref{eq:parameter_transition_arithmetic}.
The liquid-gas interface is subject to an interface heat flux $\interfaceHeatSource$ that comprises the laser heat source $\qLaserSharp$ \eqref{eq:laser_gauss_sharp} and the evaporation-induced cooling $\qVaporSharp(T)$, the latter according to \eqref{eq:evaporative_heat_loss_with_specific_enthalpy_sharp} since no vapor flow is explicitly resolved in this example.
The interface heat source $\interfaceHeatSource$ is modeled as a volumetric heat flux $\volumetricHeatSource = \interfaceHeatSource\,\deltaSingleWeighted(\indicator)$ within the diffuse interface region $\DiffuseInterfaceRegion$ using the parameter-scaled CSF.
We determine the evaporation-induced cooling $\qVaporSharp(T)$ based on the local value of the temperature according to \eqref{eq:evaporative_heat_loss_continuous_evaluation}, corresponding to the CE method.
While this assumption may result in a less accurate solution, as discussed in Section~\ref{sec:onedimEvaporation}, it is instrumental in avoiding the high computational costs in 3D associated with the extension algorithm.
Our immediate focus will be optimizing the extension algorithm's performance to use it in large-scale 3D simulations in the future.
For capturing the interface between the liquid and the solid domain, we use a regularized level set function $-1 \leq \levelset \leq 1$ \cite{kronbichler2018fast,olsson2007conservative}, the initial condition of which is determined from the initial signed distance to the interface midplane according to
\begin{align} \label{eq:initial_levelset}
	\levelset(\Bx) = \tanh\left( \frac{3\,\distance(\Bx)}{\interfaceThickness} \right)
	\quad\text{ in }\Omega\times\{t=0\}
	\text{.}
\end{align}
Via the solution of the advection equation
\begin{align} \label{eq:advection_equation}
	\fracPartial{\levelset}{t} + \Bu\cdot\nabla\levelset = 0
	\quad\text{ in }\Omega\times[0,t]
	\text{,}
\end{align}
the temporal evolution of the level set is determined.
To maintain the regularized characteristic shape of the level set function, reinitialization, according to \cite{olsson2007conservative}, is also performed.
The signed distance to the interface midplane is obtained from the regularized level set function \eqref{eq:initial_levelset} by:
\begin{align} \label{eq:signed_distance_from_levelset}
	\distance(\levelset) = \frac{\interfaceThickness}{6} \log\left( \frac{1 + \levelset}{1 - \levelset} \right)
	\text{.}
\end{align}
The latter can be used to calculate the indicator function according to \eqref{eq:indicator}.
The flow velocity $\Bu$ and pressure $p$ of the melt pool are governed by the incompressible Navier--Stokes equation, composed of the continuity equation and momentum balance equation
\begin{align}
	\nabla\cdot\Bu &= 0
	&\text{in }\Omega\times[0,t]
	\label{eq:continuity_equation} \\
	\rhoEff \left(\fracPartial{\Bu}{t} + \left( \Bu\cdot\nabla\right) \Bu \right) &= -\nabla p + \muEff\Delta\Bu + \fRecoilPressure + \fSurfaceTension + \fDarcy
	&\text{in }\Omega\times[0,t]
	\label{eq:momentum_equation}
\end{align}
with the viscosity $\mu$, the recoil pressure force $\fRecoilPressureSharp$, the surface tension $\fSurfaceTension$, and the Darcy damping term $\fDarcy$.
Here, the density $\rhoEff$ and the viscosity $\muEff$ are interpolated individually across the liquid-gas interface according to \eqref{eq:parameter_transition_arithmetic}.
The recoil pressure force $\fRecoilPressureSharp$ modeled as a continuum surface force with the density-scaled CSF model and the CE method according to \eqref{eq:recoil_pressure_continuous_evaluation}.
Similarly, the surface tension is modeled as a density-scaled continuum surface force in the sense of \cite{kothe1996volume} according to
\begin{align}
	\fSurfaceTension(\levelset) = \sigma \curvatureLG(\levelset) \normalLG(\levelset) \deltaDensity(\indicator(\levelset))
\end{align}
with the surface tension coefficient $\sigma$ and the curvature $\curvatureLG(\levelset)$ of the liquid-gas interface.
To model the melting and solidification of the metal, we employ a temperature-dependent Darcy damping force \cite{voller1987fixed} to the liquid domain that inhibits motion where the temperature is below the melting point.
By inhibiting velocities in the fluid, a rigid solid domain is modeled \cite{courtois2014complete}.
The Darcy damping term $\fDarcy$ in \eqref{eq:momentum_equation} is determined according to
\begin{align} \label{eq:darcy_damping_term}
	\fDarcy = -\darcyMorthology\left( \frac{1 - \liquidFraction^{2}}{\liquidFraction^{3} + \darcyDivZero} \right) \Bu
	\quad \text{with} \quad
	\liquidFraction = 1 - \solidFraction
\end{align}
with the liquid fraction $\liquidFraction$, the parameter $\darcyMorthology$ that depends on the mushy zone morphology, and a parameter $\darcyDivZero$ to avoid division by zero.
The solid fraction $\solidFraction$ is determined by
\begin{align} \label{eq:solid_fraction}
	\solidFraction(T,\indicator) = \begin{cases}
		\indicator &\text{for} \quad T \leq \Ts \\
		\indicator\,\frac{\Tl - T}{\Tl - \Ts} &\text{for} \quad \Ts < T < \Tl \\
		0 &\text{otherwise}
	\end{cases}
\end{align}
with the liquidus temperature $\Tl$ and the solidus temperature $\Ts$.
Typical material parameter values for \tiSixFour are employed, listed in Table~\ref{tab:param_ti64}.
The values for $\Tl$, $\darcyMorthology$, and $\darcyDivZero$ are chosen to achieve a sufficiently smooth transition between the mobile liquid phase and the rigid solid phase.

We consider the 3D cuboid domain ${\Omega = \{\Bx \,|\, x,y \in [-a,a], z \in [-b,b]\}}$ with the length parameters $a = \SI{400}{\mu m}$ and $b = \SI{300}{\mu m}$.
The initial metal-gas interface coincides with the $xy$-plane, characterized by the initial signed distance of $d(\Bx) = -z$.
The fluid is initially at rest ($\Bu_{0} = \Bzero$, $p_{0} = 0$) and the initial temperature is uniform at $T_{0} = \SI{500}{K}$ \eqref{eq:heat_equation_initial_condition}.
At the bottom ($z = -b$) and top ($z = b$) boundaries, no-slip conditions for the incompressible Navier--Stokes equations are assumed, and the temperature is prescribed to $\bar{T} = \SI{500}{K}$ by a Dirichlet boundary condition \eqref{eq:heat_equation_dirichlet}.
Along the vertical boundaries, we assume slip conditions for incompressible Navier--Stokes equations and adiabatic boundary conditions for the heat equation.
The metal surface is exposed to a spatially fixed laser heat source with a Gaussian profile according to \eqref{eq:laser_gauss_sharp} considering a laser power of $\laserPower = \SI{156}{W}$, a laser radius $\laserRadius = \SI{70}{\mu m}$, and the laser beam direction $\laserDirection$ corresponding to the negative $z$-direction with a characteristic point along the laser beam axis $\laserPosition$ at the origin.

To solve the governing equations \eqref{eq:heat_equation}, \eqref{eq:advection_equation}, \eqref{eq:continuity_equation}, \eqref{eq:momentum_equation}, we employ spatial discretization by the FEM with a Cartesian mesh and linear shape functions for the level-set, the pressure and the temperature.
To ensure inf-sup stability, we use quadratic shape functions for the velocity field.
Using adaptive mesh refinement, the grid with a base element size of $h_{\*{max}} = \SI{23.1}{\mu m}$ is locally refined to a $h_{\*{min}} = \SI{2.89}{\mu m}$ in the vicinity of the diffuse interface region.
Employing an interface thickness of $\interfaceThickness = \SI{30}{\mu m}$ results in about $\numberOfElementsInInterface \approx 10$ elements across the interface.
We employ operator splitting, considering a weakly partitioned solution scheme for solving the coupled system of equations.
For the individual subproblems, implicit time-stepping schemes are considered.
A detailed description of the employed numerical two-phase flow framework and solution strategy is given in \cite{kronbichler2018fast,schreter2024consistent}.
For the present example, a constant time step size of $\Delta t = \SI{e-8}{s}$ is considered.

\begin{figure}[tbp!]
	\vspace{0.75cm}
	\subfloat[$t = \SI{0.04}{ms}$]{
		\includegraphics[width=0.23\textwidth]{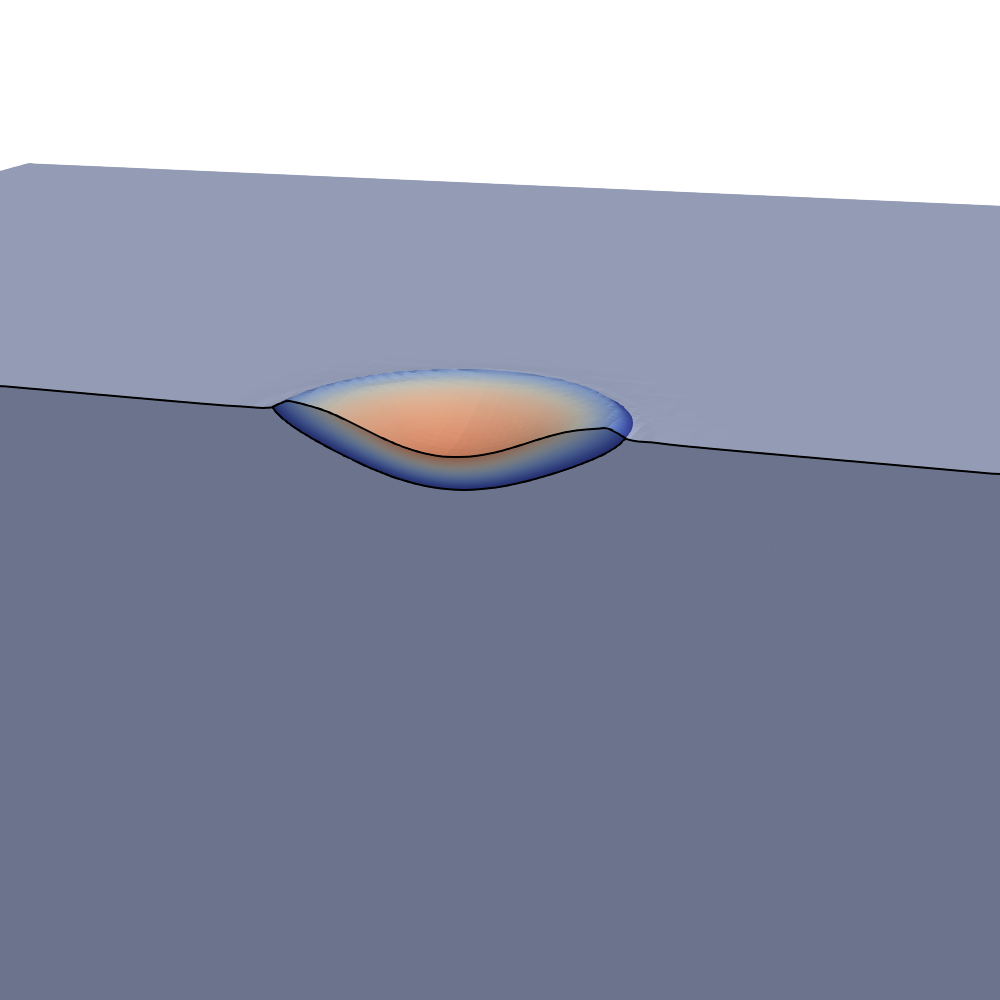}
	}
	\subfloat[$t = \SI{0.08}{ms}$]{
		\includegraphics[width=0.23\textwidth]{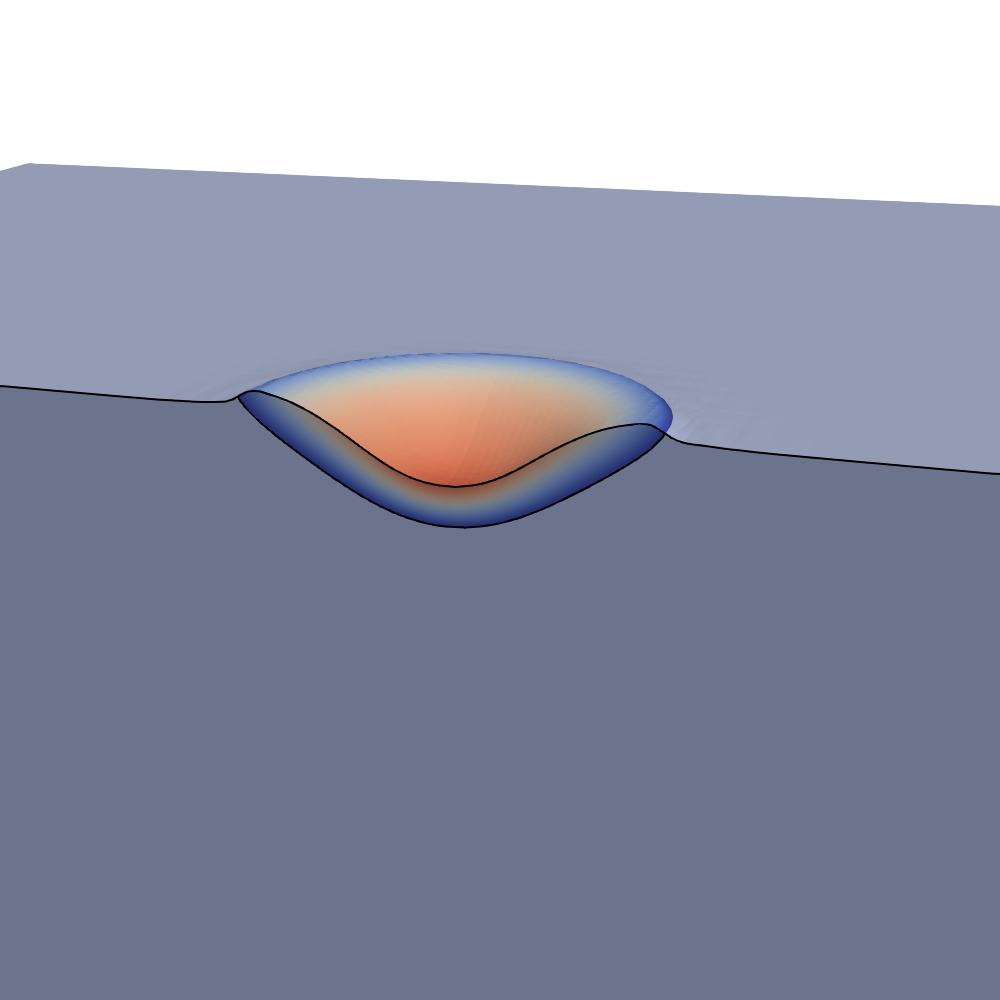}
	}
	\subfloat[$t = \SI{0.12}{ms}$]{
		\includegraphics[width=0.23\textwidth]{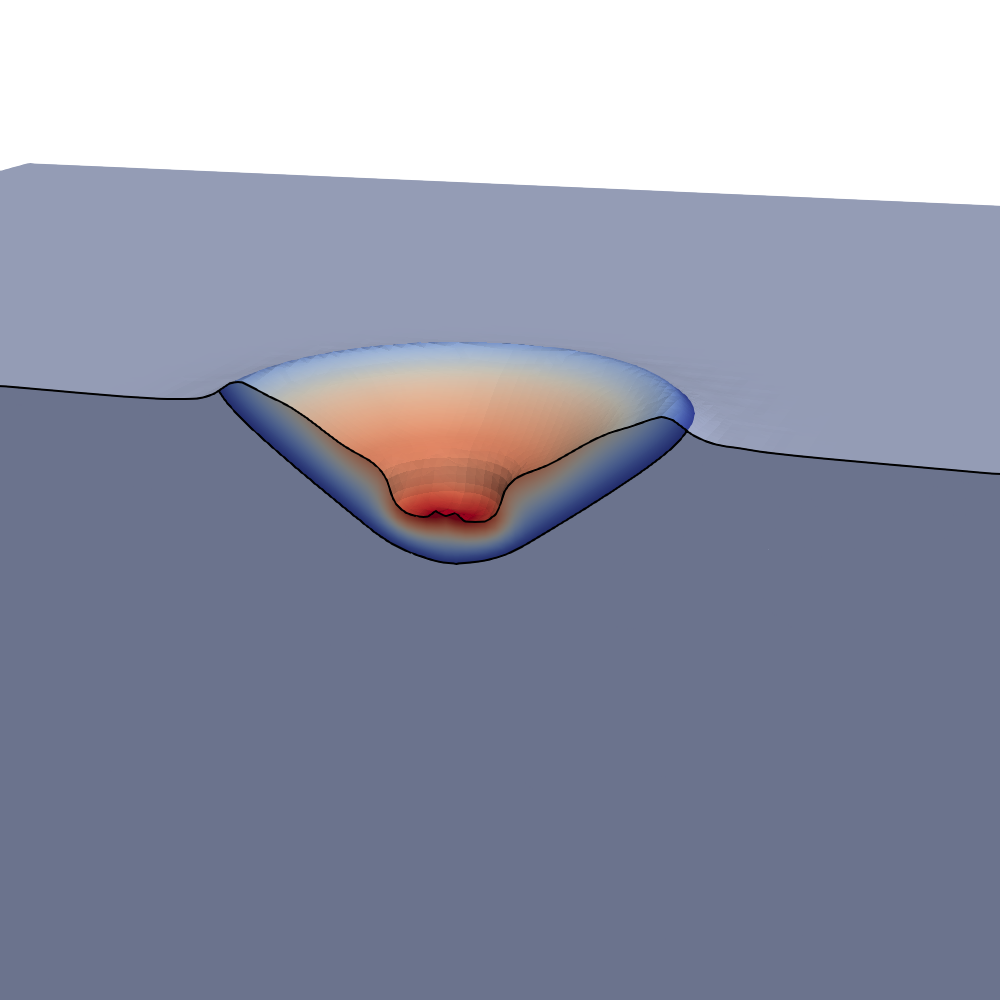}
	}
	\subfloat[$t = \SI{0.16}{ms}$]{
		\includegraphics[width=0.23\textwidth]{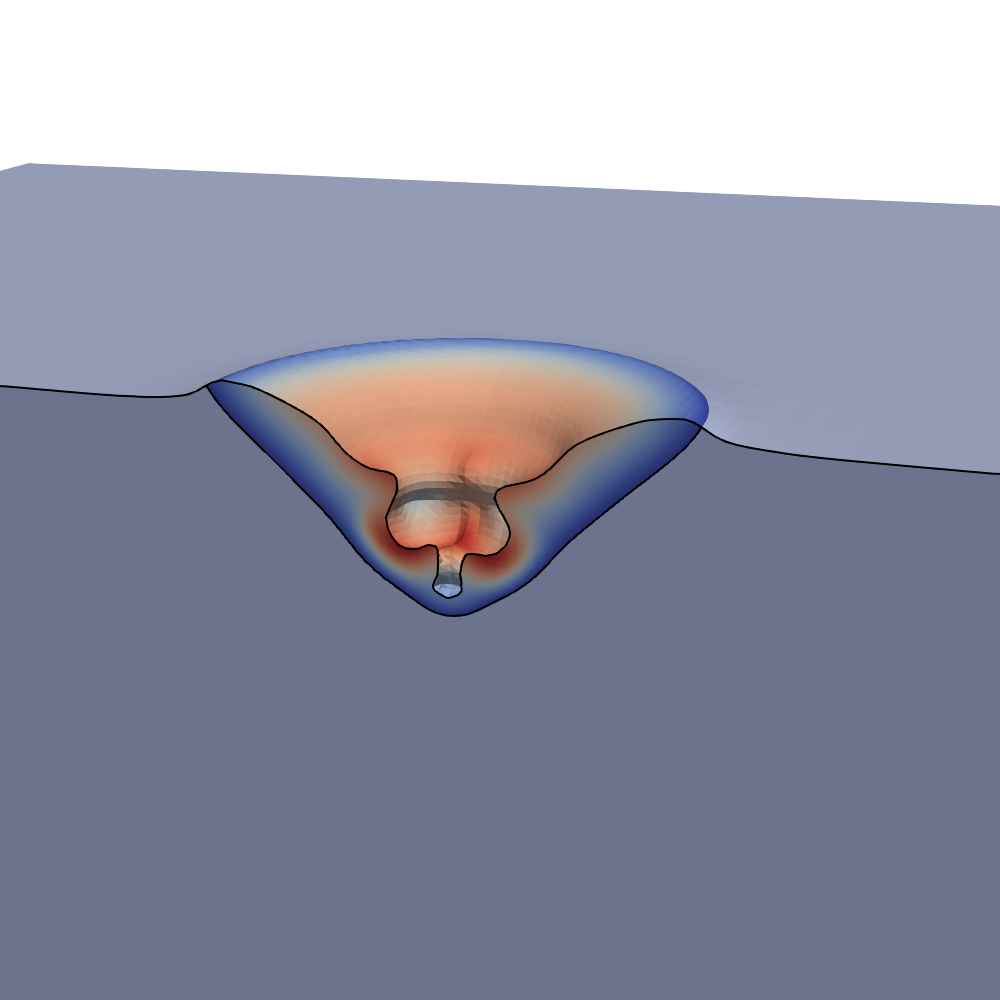}
	}
	\\
	\subfloat[$t = \SI{0.24}{ms}$]{
		\includegraphics[width=0.23\textwidth]{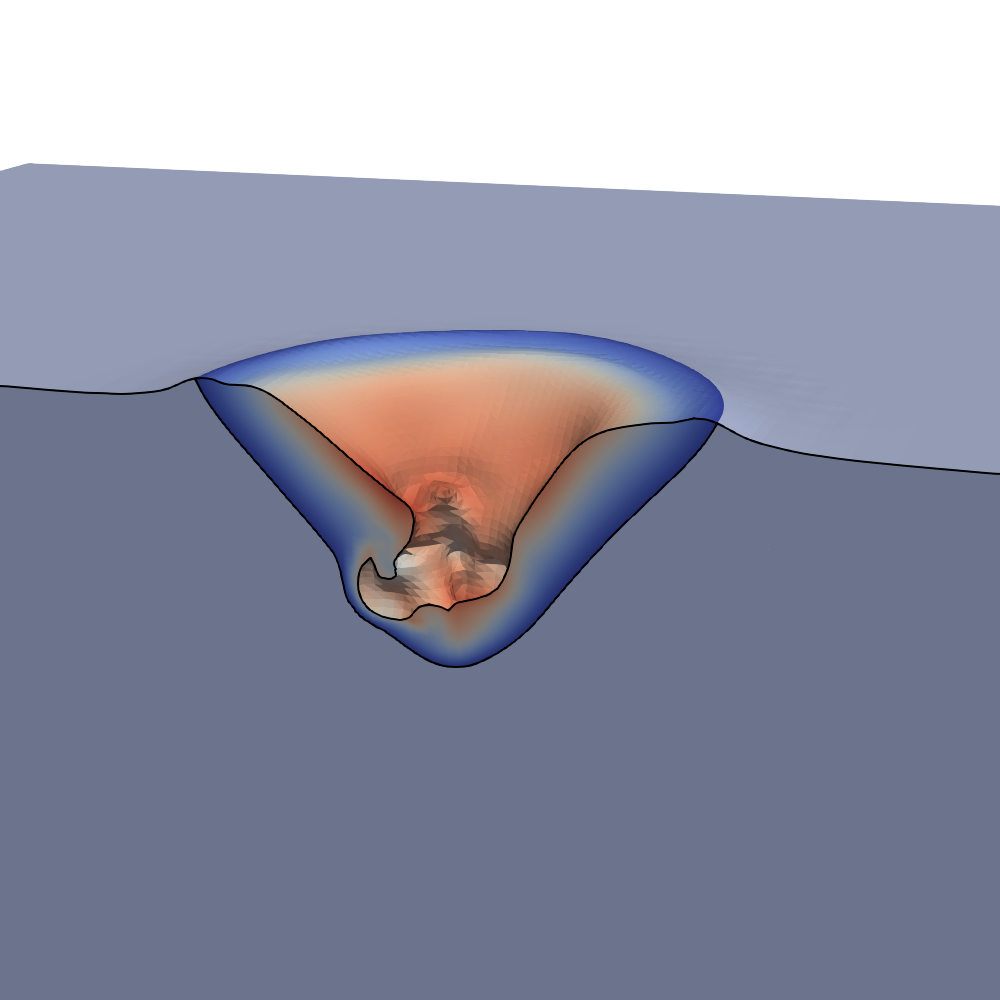}
	}
	\subfloat[$t = \SI{0.32}{ms}$]{
		\includegraphics[width=0.23\textwidth]{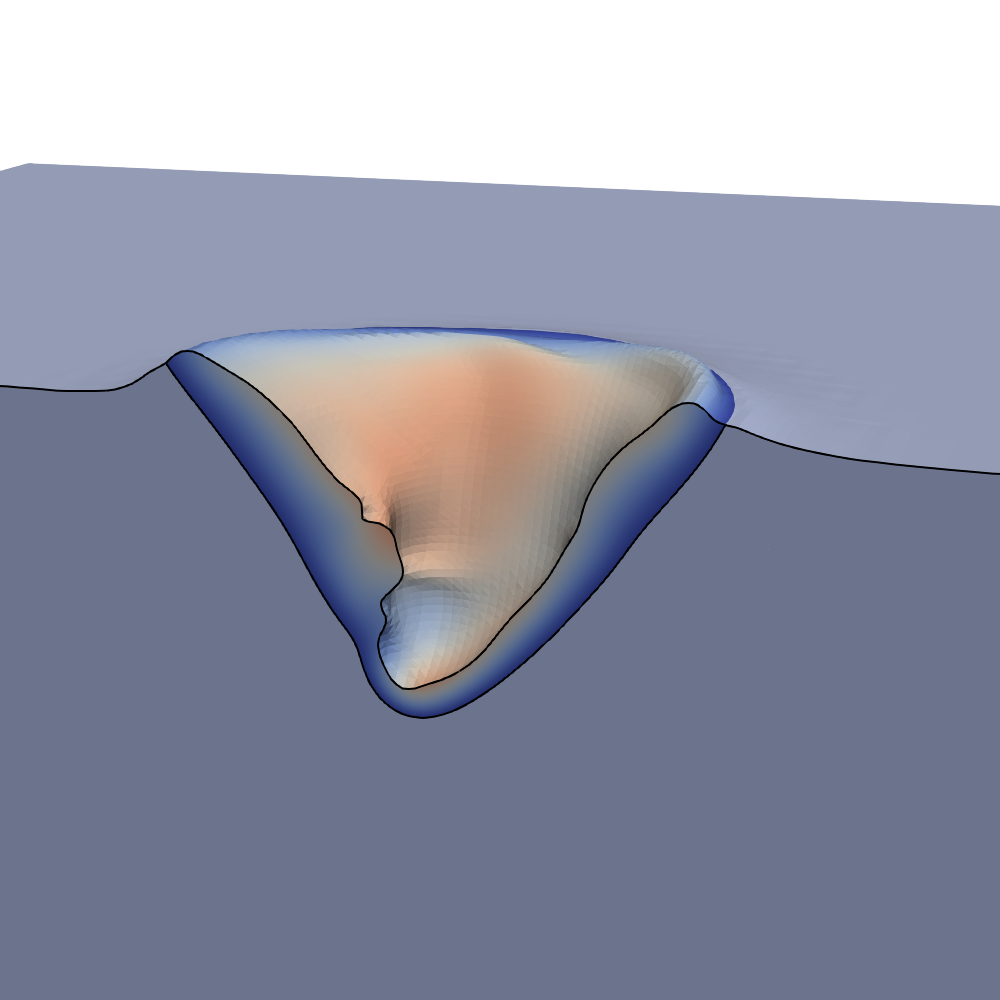}
	}
	\subfloat[$t = \SI{0.40}{ms}$]{
		\includegraphics[width=0.23\textwidth]{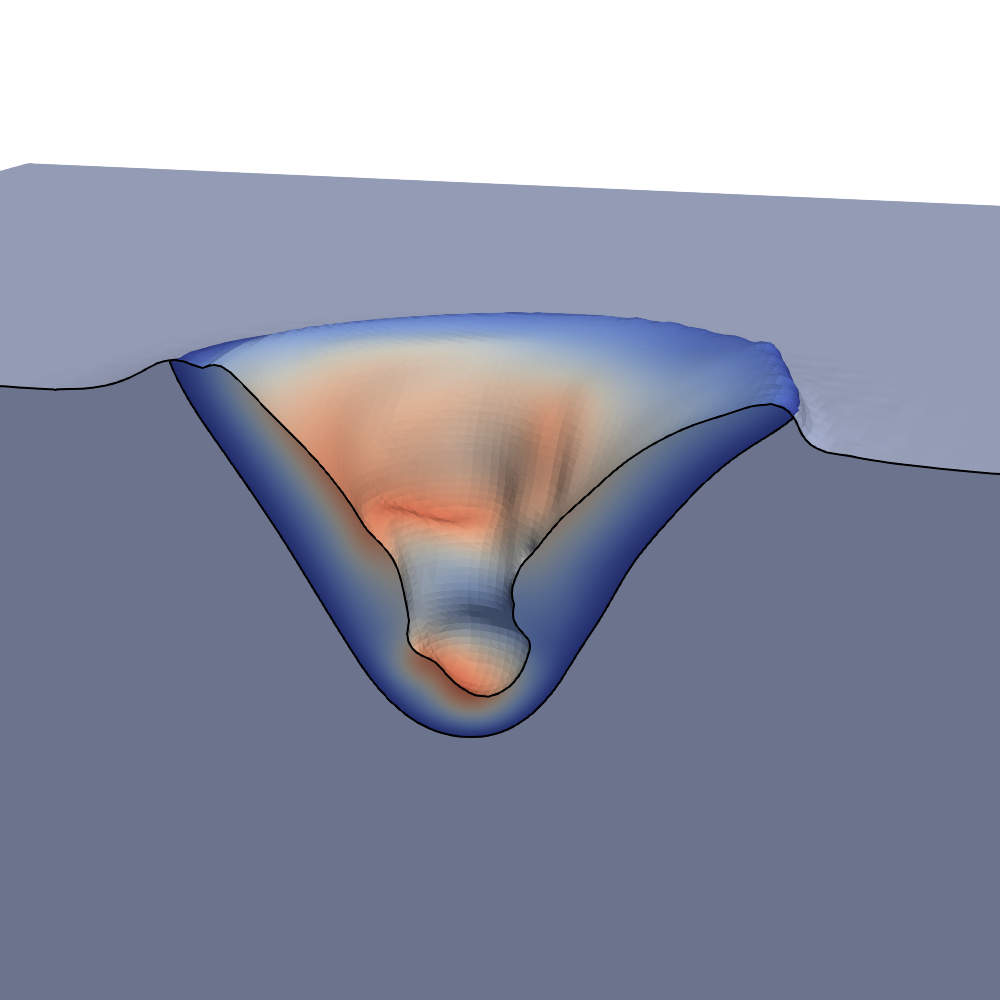}
	}
	\subfloat[$t = \SI{0.48}{ms}$]{
		\includegraphics[width=0.23\textwidth]{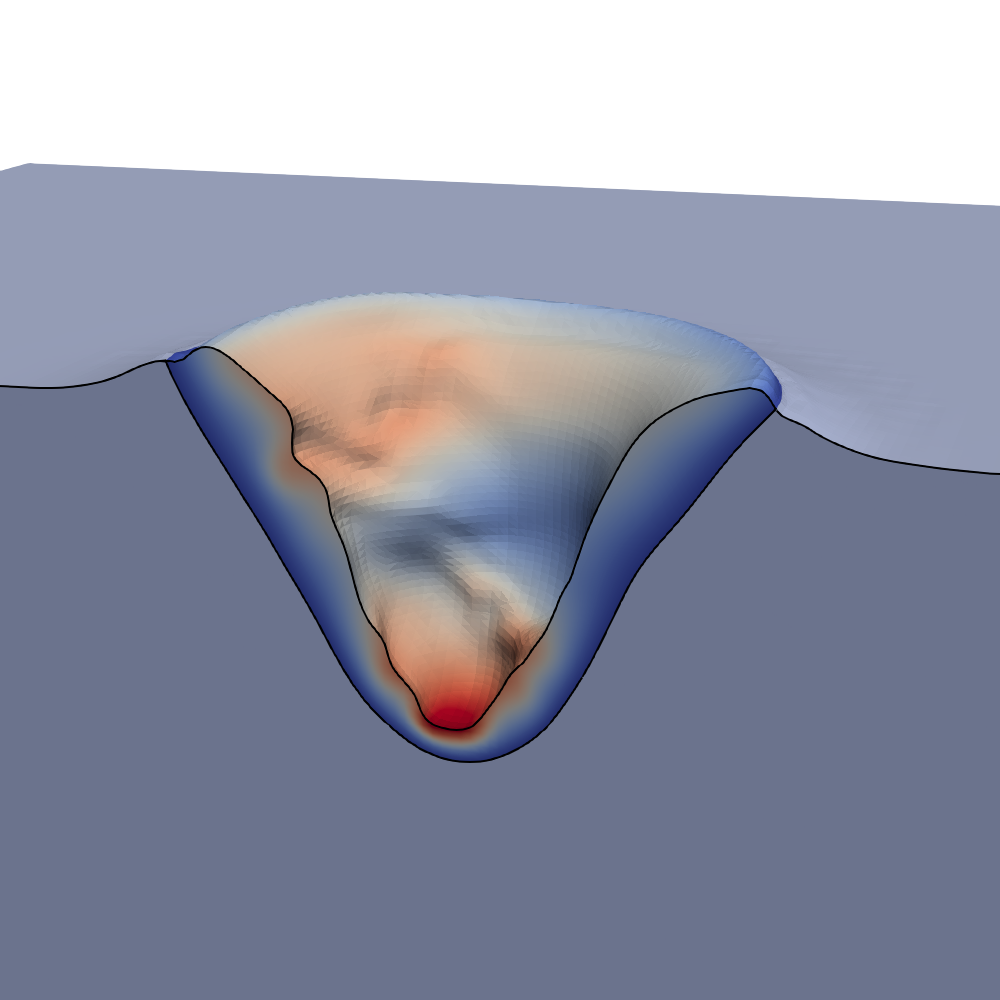}
	}
	\tikz[overlay,remember picture] \node[anchor=east] at (-3.3cm, 7.5cm)
		{\colorbarHorizontal{5cm}{1933}{3500}{\footnotesize temperature (liquid) (K)}};

	\caption{Sectional view of the fully coupled 3D melt pool thermo-hydrodynamics simulation of stationary laser-induced heating of a bare \tiSixFour{} plate:
	time-series illustrating the melt pool shape with the temperature field of the liquid domain.
	}
	\label{fig:fully_coupled_3D_simulation_snapshots}
\end{figure}

Fig.~\ref{fig:fully_coupled_3D_simulation_snapshots} presents sectional view snapshots from the simulation, depicting the temperature distribution in the liquid domain at different time steps.
Typical characteristics for \PBFAM processing in the keyhole mode can be observed:
Upon attaining the boiling temperature, the evaporation-induced recoil pressure increases, and a stable vapor depression forms.
As the vapor depression grows, instabilities start to form due to fluctuations in the recoil pressure and in conjunction with the surface tension.
The oscillations increase until they become unstable, and the melt pool transitions to highly dynamic and chaotic motion.
In this simulation, the melt pool becomes unstable at approx.~$t = \SI{0.12}{ms}$.
This behavior is also seen in experiments by Cunningham et al.~\cite{cunningham2019keyhole}.
The results indicate that the present model is capable of replicating important characteristics of melt pool behavior.

It should be noted that the results of this simulation do not claim high accuracy.
Extrapolating the results from Section~\ref{sec:twodimFixedMeltPool}, the discretization and interface thickness chosen for the present 3D simulation suggest that the recoil pressure is likely to be very inaccurate.
A further critical aspect is related to the simplified modeling of the laser energy absorption (see Table~\ref{tab:param_ti64}).
For the laser absorptivity, we considered $\absorptivity = 0.35$, which Khairallah et al.~\cite{khairallah2016laser} calibrated on a similar computational melt pool model to show the best agreement with an experiment.
However, calibrating the absorptivity may contribute to compensating for the inaccuracy of diffuse interface models.
In addition, this simplified absorption model does not replace the realistic consideration of the overall laser impact, e.g., via a ray-tracing model, which is particularly relevant in the keyhole regime and a pending feature of our model.

In this section, a highly dynamic melt pool simulation based on the experimental setup by Cunningham et al.~\cite{cunningham2019keyhole} is used to demonstrate the robustness and applicability of the parameter-scaled CSF model to a challenging, practically relevant problem type.
It should be noted that when we applied the classical CSF approach to the same problem, we could not achieve convergence of the involved nonlinear solvers for the Navier-Stokes/heat transfer equations, which may be due to the high gradients induced by the classical CSF approach as shown in Section~\ref{sec:classicalCSF}.
Using the parameter-scaled CSF, the simulation reproduces key characteristics of the behavior in the experiment.

\section{Discussion: Suitability of diffuse interface melt pool models}
\label{sec:discussion}

In this study, we address a fundamental question in computational melt pool modeling: the accuracy of diffuse interface approaches for reliable predictions of melt pool dynamics.
Diffuse models are widely favored for their inherent robustness and straightforward mathematical formulation but have not been thoroughly evaluated for their inherent modeling errors and convergence properties (see \cite{cook2020simulation}) to the best of our knowledge.
Particularly, their accuracy in capturing critical phenomena in \PBFAM melt pool dynamics related to the interface temperature, such as evaporation-induced effects, has been underexplored.
To this end, we systematically investigated the accuracy of continuum surface flux (CSF) methods for capturing interface effects in two-phase heat transfer modeling, focusing on realistic material property ratios representative of \PBFAM.
In addition, we proposed novel approaches to enhance the accuracy of such models.

In Sections~\ref{sec:classicalCSF}-\ref{sec:twodimFixedMeltPool}, we investigated the accuracy of CSF methods with respect to the chosen discretization parameters -- interface thickness and mesh resolution -- for two-phase heat transfer modeling on reduced-order benchmark examples representative of \PBFAM.
Compared to classical CSF modeling in the sense of Brackbill et al.~\cite{brackbill1992continuum}, our proposed parameter-scaled CSF approach improves the temperature prediction accuracy by one order of magnitude for given discretization parameters.
However, accurate modeling of interface dynamics in \PBFAM melt pool simulations necessitates precise consideration of interface temperature-dependent effects, including evaporation-induced cooling and recoil pressure.
This accuracy is directly related to the temperature in the interface region, which is particularly prone to inherent modeling errors associated with diffuse interface approaches.
Our analysis demonstrated that errors in the recoil pressure -- which is one of the major driving forces for melt pool dynamics -- are about one order of magnitude higher than errors in temperature due to the underlying exponential dependence.
Therefore, to achieve sufficiently accurate predictions of the overall melt pool dynamics for application scenarios with realistic temperature fields and material parameters, it is imperative to employ very small interface thickness values and correspondingly fine mesh resolutions within the interface region.
Consequently, more efficient implementations and the use of high-performance computing infrastructures at a larger scale are required.

There exists a large variety of \PBFAM melt pool models, e.g., based on finite difference, finite volume, finite element, lattice Boltzmann, or meshfree discretizations, which mostly utilize diffuse interface approaches.
Often, an insufficient agreement of such models with experimental measurements is reported, as thoroughly analyzed, e.g., in \cite{andreotta2017finite,khairallah2016laser,ross2022volumetric,zhu2021mixed}.
In many cases, this problem is addressed by fitting model parameters such as the laser absorptivity to match experimental measurements.
In contrast, the results of the present study suggest that the extreme temperature gradients close to the melt pool surface, as typical for \PBFAM, in combination with an insufficient resolution of the diffuse interface domain, might also be a potential explanation for this shortcoming.
While critical model parameters such as laser absorptivity are often unknown a priori, thus requiring careful calibration, such a calibration procedure can, of course, not compensate for potential discretization errors due to insufficient interface resolution in general scenarios.
Thus, for future research, detailed investigations on the accuracy of the different existing types of diffuse and sharp interface approaches, when applied to \PBFAM melt pool modeling, are recommended.

\section{Conclusions}
\label{sec:conclusions}

Many existing computational models for studying melt pool dynamics in \PBFAM rely on a diffuse interface description of the underlying thermo-hydrodynamic two-phase problem.
In such models, the accurate modeling of the temperature is a prerequisite for the realistic prediction of melt pool dynamics as the governing forces, such as evaporation-induced cooling and recoil pressure, are exponentially related to the interface temperature.
Thus, quantifying the inherent modeling error and the convergence properties of the diffuse interface approach is needed.
For this purpose, we performed a comprehensive study of thermal two-phase problems representing \PBFAM in a diffuse finite element framework.
We considered sharp-interface reference solutions to measure the error in terms of the temperature field and the resulting evaporation-induced recoil pressure.

We demonstrated that when a classical CSF approach is applied, along with typical interface thicknesses and discretizations, the extreme temperature gradients beneath the melt pool surface, as induced by the localized energy input in \PBFAM, combined with the high ratios of thermal conductivity $(\sim 10^{3})$ and volume-specific heat capacity $(\sim 10^{5})$ between metal and ambient gas, lead to significant errors in the interface temperature.
As a promising alternative, we propose a novel parameter-scaled CSF approach to obtain a smoother temperature rate in the diffuse interface region, thus significantly increasing the solution accuracy.
It has been shown that the criterion for the required interface thickness to predict the temperature field with a given level of accuracy is less restrictive by at least one order of magnitude for the proposed parameter-scaled approach compared to classical CSF.
For 3D problems, the number of discretization points within the diffuse interface region scales quadratically with the interface thickness, given a constant resolution across the interface, which underlines the relevance of this result in terms of significantly reducing the computational cost.

Additionally, we showed that evaluating the temperature at the interface midplane for the computation of temperature-dependent diffuse interface fluxes, instead of using local values across the interface thickness, yields a more accurate interface temperature and, consequently, a more accurate recoil pressure.

Notably, our findings extend beyond the pure thermal problem as we showcased the general applicability of the parameter-scaled CSF to a 3D simulation of stationary laser melting considering the fully coupled thermo-hydrodynamic multi-phase problem, including phase change.

Finally, our conclusion that the extreme temperature gradients in \PBFAM in combination with an insufficient resolution of the diffuse interface domain might lead to significant modeling errors of the overall melt pool dynamics is expected to be highly relevant also for other types of diffuse interface approaches.
Thus, detailed investigations on the accuracy of the different existing types of diffuse and sharp interface approaches, when applied to \PBFAM melt pool modeling, are recommended for future research.

\section*{Declarations}

\bmhead{Availability of data and materials}
The research code, numerical results, and digital data obtained in this project are held on deployed servers that are backed up.
The datasets used and/or analyzed during the current study are available from the corresponding author upon reasonable request.

\bmhead{Competing interests}
The authors declare that they have no competing interests.

\bmhead{Funding}
Magdalena Schreter-Fleischhacker received funding by the Austrian Science Fund (FWF) Schrödinger Fellowship (project number: J4577).
Christoph Meier gratefully acknowledges the financial support from the European
Research Council through the ERC Starting Grant \emph{ExcelAM} (project number: 101117579).

\bmhead{Author contributions}
NM and MS contributed to the derivation of model equations and to the specific code implementation.
NM was responsible for the numerical studies.
PM supported the implementation.
In addition, PM and MK contributed general-purpose functionality to this project via the \texttt{deal.II} library and the \texttt{adaflo} project.
MS, CM, and WAW worked out the general conception of the proposed modeling approach.
All authors participated in writing	and discussion of the manuscript.

\bmhead{Acknowledgements}
Not applicable

\begin{appendices}

\section{Continuum surface flux model: temperature rate}
\label{app:appendix_CSF_temperature_rate}

Let us consider the heat equation \eqref{eq:heat_equation} without heat convection
\begin{align}
	\underbrace{\interp{\left(\rho\cp\right)}}_{\volumetricCapacityEff} \fracPartial{T}{t} = \nabla \left(\interp{\conductivity}\,\nabla T\right) + \volumetricHeatSource
	\quad\text{ in }\Omega\times[0,t]
	\text{.}
\end{align}
The volumetric heat flux $\volumetricHeatSource$ models an interface heat flux $\volumetricHeatSource = \interfaceHeatSource\,\deltaI(\indicator)$ with a not yet specified delta function $\deltaI(\indicator)$ as a result of the CSF modeling.
We assume that the volumetric heat flux $\volumetricHeatSource$ dominates the problem, which results in a low Fourier number $\Fo$ \eqref{eq:fourier_number}.
This allows us to neglect the conductive term if only the short-term behavior of a problem shall be studied:
\begin{align} \label{eq:heat_equation_reduced}
	\volumetricCapacityEff(\indicator) \fracPartial{T}{t} = \interfaceHeatSource\,\deltaI(\indicator)
	\text{.}
\end{align}
Note that the effective volume-specific heat capacity $\volumetricCapacityEff(\indicator)$ is interpolated between the two phases using the indicator $\indicator$ and, for example, the arithmetic mean interpolation \eqref{eq:parameter_transition_arithmetic}.
To obtain the temperature rate $\fracPartial{T}{t}$, we divide by the effective volume-specific heat capacity $\volumetricCapacityEff(\indicator)$ because the thermal mass is proportional to it.
\begin{align} \label{eq:temperatureRateA}
	\fracPartial{T}{t} = \frac{\interfaceHeatSource\,\deltaI(\indicator)}{\volumetricCapacityEff(\indicator)}
\end{align}
As can be seen from \eqref{eq:temperatureRateA}, the temperature rate $\fracPartial{T}{t}$ is inversely proportional to the volume-specific heat capacity.
Since the volume-specific heat capacity is an interpolated quantity within the diffuse interface, using the symmetric delta function, the shape of the profile of the temperature rate $\fracPartial{T}{t}$ is skewed, see the bottom right panel of Fig.~\ref{fig:interface_heat_source_by_capacity_symdelta}.
We can exploit the freedom in choosing the delta function $\deltaI$ to counteract the skew shape due to the diffuse material parameters.
Setting the norm of the indicator gradient $\deltaFunction(\indicator)$ \eqref{eq:norm_of_indicator_gradient} as the goal shape of the temperature rate, we choose the sought-after delta function proportional to the norm of the indicator gradient $\deltaFunction(\indicator)$ and the effective volume-specific heat capacity $\volumetricCapacityEff(\indicator)$, i.e., the thermal mass:
\begin{align} \label{eq:deltaCv}
	\deltaI = \deltaFunction(\indicator)\,\volumetricCapacityEff(\indicator) \, c_{\*{corr}}
	\text{.}
\end{align}
The correction factor $c_{\*{corr}}$ has to be chosen such that the delta function satisfies the condition of identity \eqref{eq:delta_identity}.
By inserting \eqref{eq:deltaCv} in \eqref{eq:temperatureRateA}, we obtain
\begin{align}
	\fracPartial{T}{t}
	= \frac{\interfaceHeatSource\,\deltaFunction(\indicator)\,\volumetricCapacityEff(\indicator)\,c_{\*{corr}}}{\volumetricCapacityEff(\indicator)}
	= \interfaceHeatSource\,c_{\*{corr}}\,\deltaFunction(\indicator)
	\text{,}
\end{align}
which meets the goal shape of the temperature rate by the norm of the indicator gradient $\deltaFunction(\indicator)$ scaled with a constant pre-factor of $\interfaceHeatSource\,c_{\*{corr}}$.
Although the obtained temperature rate has the goal shape, its magnitude now depends on the correction factor $c_{\*{corr}}$ in addition to the interface heat flux $\interfaceHeatSource$.
The correction factor depends on the interpolation type of the volume-specific heat capacity $\volumetricCapacityEff(\indicator)$, and in Section~\ref{sec:parameterScaledCSFdelta}, the formulations for some interpolation types are given.
In the bottom right panel of Fig.~\ref{fig:interface_heat_source_by_capacity_asymdelta}, it is shown that for a variety of interpolation types, the temperature rate is well distributed, but the magnitude varies significantly.
However, the change in temperature rate due to a change in the interpolation type does not change the heat flow rate of the interface heat flux, which is shown in Appendix~\ref{app:appendix_CSF_energy_rate}.

\section{Continuum surface flux model: energy rate}
\label{app:appendix_CSF_energy_rate}

Consider a two-phase domain $\Omega = \OmegaG \cup \OmegaL$ in which the heat transfer is governed by the heat equation \eqref{eq:heat_equation}.
An interface heat flux $\interfaceHeatSource$ is imposed at the interface $\GammaLG$ between the two phases.
Using the parameter-scaled CSF method, the interface heat flux is modeled as a volumetric heat flux $\volumetricHeatSource = \interfaceHeatSource\,\deltaI(\indicator)$ in the diffuse interface region $\DiffuseInterfaceRegion \in \Omega$ characterized by the interface thickness $\interfaceThickness$ around the interface midplane $\GammaLG$.
Here, $\deltaI$ is the appropriate delta function for the chosen interpolation of the effective volume-specific heat capacity $\volumetricCapacityEff$, see Section~\ref{sec:parameterScaledCSFdelta}.
The total input heat flow rate $\dot{Q}$ resulting from an interface heat flux $\interfaceHeatSource$ is:
\begin{align} \label{eq:energy_rate_input_global}
	\dot{Q}
	= \int_{\Omega} \volumetricHeatSource \,\diffd\Omega
	= \int_{\Omega} \interfaceHeatSource\,\deltaI(\indicator) \,\diffd\Omega
	\text{.}
\end{align}
All parameter-scaled delta functions $\deltaI(\indicator)$ presented in Section~\ref{sec:parameterScaledCSFdelta} have support only in the diffuse interface region $\DiffuseInterfaceRegion$, because they are proportional to the norm of the indicator gradient $\deltaFunction(\indicator)$ \eqref{eq:norm_of_indicator_gradient}.
Thus, the delta function is zero outside the diffuse interface region $\DiffuseInterfaceRegion$, and the domain integral in \eqref{eq:energy_rate_input_global} can be rewritten as an integral over the diffuse interface
\begin{align} \label{eq:energy_rate_input_interface}
	\dot{Q}
	= \int_{\DiffuseInterfaceRegion} \interfaceHeatSource\,\deltaI(\indicator) \,\diffd\Omega
	= \int_{\GammaLG} \int_{-\frac{\interfaceThickness}{2}}^{\frac{\interfaceThickness}{2}} \interfaceHeatSource\,\deltaI(\indicator(\distance)) \,\diffd\distance \,\diffd\Gamma\,
	\text{.}
\end{align}
As the interface heat flux $\interfaceHeatSource$ is constant across the interface thickness and all delta functions have to satisfy \eqref{eq:delta_identity}, \eqref{eq:energy_rate_input_interface} can be rewritten as
\begin{align}
	\dot{Q}
	= \int_{\GammaLG} \interfaceHeatSource \int_{-\frac{\interfaceThickness}{2}}^{\frac{\interfaceThickness}{2}} \deltaI(\indicator(\distance)) \,\diffd\distance \,\diffd\Gamma
	= \int_{\GammaLG} \interfaceHeatSource \,\diffd\Gamma
	\text{,}
\end{align}
which results in the same expression as in a sharp interface model.
This shows that the CSF model does not alter the input heat flow rate.
\\

In Section~\ref{sec:parameterScaledCSF}, in particular in Fig.~\ref{fig:interface_heat_source_by_capacity_asymdelta} and Fig.~\ref{fig:onedim_asymdelta_dyn_stat_h_stud_consteps_1e-6_T-profiles}, we observed different temperature rate and temperature magnitudes for different interpolation types of the effective volume-specific heat capacity $\volumetricCapacityEff$ as a result from interface heating.
As shown above, the total input heat flow rate $\dot{Q}$ of the CSF interface heat flux has to be the same for all parameter-scaled delta functions $\deltaI$.
In a next step, the internal energy shall be investigated.
As discussed in Section~\ref{sec:classicalCSFthermal}, the heat conduction term in \eqref{eq:heat_equation} can be neglected in good approximation if only the short-term behavior is considered.
With this assumption, it can be concluded from the integration of \eqref{eq:heat_equation_reduced} over the spatial domain that an unchanged value of the total input heat flux rate $\dot{Q}$ leads to a rate of change in internal energy $\dot{E}_{T}$ according to
\begin{align} \label{eq:energy_rate_thermal}
	\dot{E}_{T} = \int_{\Omega} \volumetricCapacityEff(\indicator(\Bx)) \, \fracPartial{T(\Bx)}{t} \,\diffd\Omega
\end{align}
that is unchanged, i.e., independent from the chosen interface interpolation and delta functions, as well since, in this case, $\dot{E}_{T}\equiv \dot{Q}$ holds.
Within the diffuse interface region $\DiffuseInterfaceRegion \in \Omega$, the effective volume-specific heat capacity $\volumetricCapacityEff$ is determined by the choice of interpolation between the two phases, which influences its magnitude as can be seen in the top right panel of Fig.~\ref{fig:interface_heat_source_by_capacity_asymdelta}.
At a constant rate of change in internal energy $\dot{E}_{T}$, the temperature rate $\fracPartial{T}{t}$ is inversely proportional to the effective volume-specific heat capacity $\volumetricCapacityEff$.
When changing the interpolation of $\volumetricCapacityEff$, the temperature rate changes accordingly, resulting in significant differences as can be seen in the bottom right panel of Fig.~\ref{fig:interface_heat_source_by_capacity_asymdelta}.
The change in temperature rate influences the temperature magnitudes after some time passed, as can be seen in Fig.~\ref{fig:onedim_asymdelta_dyn_stat_h_stud_consteps_1e-6_T-profiles}.
However, the total energy rate, which contains the product of $\volumetricCapacityEff(\indicator(\Bx))$ and $\fracPartial{T(\Bx)}{t}$ remains unchanged in good approximation as discussed above.

\section{Investigation of the parameter-scaled CSF model on the laser-induced heating benchmark example with harmonic mean interpolation}
\label{app:parameterScaledCSFexampleHarmonic}

\begin{figure}[tbp!]
	\centering
	\subfloat[
	Instationary temperature profiles at $t=\SI{e-5}{s}$ using a CSF interface thickness of $\interfaceThickness = \SI{6}{\mu m}$ for different finite element sizes $h$ at the interface.
	]{
		\includegraphics{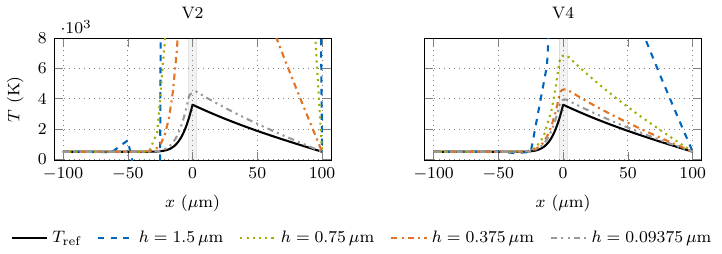}
		\label{fig:onedim_recidelta_dyn_stat_h_stud_consteps_1e-6_T-profiles}
	}
	\\
	\subfloat[
	Relative error in the instationary temperature profile at $t=\SI{e-5}{s}$ using a CSF interface thickness of $\interfaceThickness = \SI{6}{\mu m}$ for different finite element sizes $h$ at the interface.
	$\numberOfElementsInInterface$ is the number of finite elements across the interface \eqref{eq:numberOfElementsInInterface}.
	]{
		\includegraphics{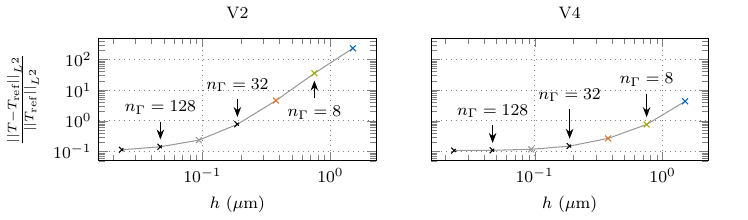}
		\label{fig:onedim_recidelta_dyn_stat_h_stud_consteps_1e-6_T-error}
	}
	\\
	\subfloat[
	Relative error in the instationary temperature profile at $t=\SI{e-5}{s}$ for different diffuse interface thicknesses $\interfaceThickness$ with ${\numberOfElementsInInterface \in\{8, 16, 32, 64, 128\}}$ finite elements across the interface.
	The interface thickness used in Fig.~\ref{fig:onedim_recidelta_dyn_stat_h_stud_consteps_1e-6_T-error} is annotated.
	]{
		\includegraphics{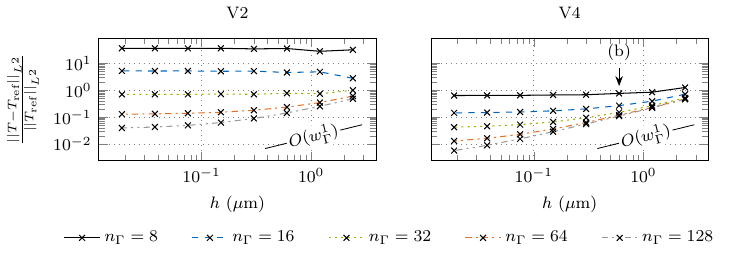}
		\label{fig:onedim_recidelta_dyn_stat_interface_width_stud_ni_128_T-error}
	}
	\caption{
		Temperature profile resulting from interface heating with an interface heat source of $\interfaceHeatSource = \SI{e10}{W\per m^2}$ using the parameter-scaled CSF approaches V2 and V4, described in Table~\ref{tab:parameter_scaled_CSF_cases}.
		The reference temperature profile $\Tref$ is determined using a sharp interface approach.
	}
	\label{fig:onedim_recidelta_studies}
\end{figure}

This section evaluates the performance of the parameter-scaled CSF for modeling interface heat fluxes for the cases V2 and V4 listed in Table~\ref{tab:parameter_scaled_CSF_cases}, which involve harmonic mean interpolation \eqref{eq:parameter_transition_harmonic}.
The same benchmark example and numerical setup are employed for cases V1 and V3 and described in Section~\ref{sec:parameterScaledCSFexample}.
Fig.~\ref{fig:onedim_recidelta_studies} shows the instationary results at $t = \SI{e-5}{s}$ with V2 in the left column and V4 in the right column.
In Fig.~\ref{fig:onedim_recidelta_dyn_stat_h_stud_consteps_1e-6_T-profiles}, the temperature profiles are shown at a constant interface thickness of $\interfaceThickness = \SI{6}{\mu m}$ and for different discretizations of the interface region.
Both variants significantly overestimate the temperature at the given interface thickness, especially for insufficient discretization.
The $L^{2}$-norm of the relative temperature error to the sharp reference solution for different element sizes $h$ in the interface, including the values shown in Fig.~\ref{fig:onedim_recidelta_dyn_stat_h_stud_consteps_1e-6_T-profiles}, at a constant interface thickness of $\interfaceThickness = \SI{6}{\mu m}$ is shown in Fig.~\ref{fig:onedim_recidelta_dyn_stat_h_stud_consteps_1e-6_T-error}.
The number of finite elements across the interface $\numberOfElementsInInterface$ \eqref{eq:numberOfElementsInInterface} is annotated.
For V4, when increasing $\numberOfElementsInInterface$ from \num{256} to \num{512}, the change in the relative temperature error is less than 1\%.
For V2, the change in relative temperature error is always larger than 1\% for the investigated discretizations.
In Fig.~\ref{fig:onedim_recidelta_dyn_stat_interface_width_stud_ni_128_T-error}, the relative temperature error is shown over the interface thickness $\interfaceThickness$ for different resolutions of the interface.
The convergence behavior of the relative temperature error with respect to the interface thickness tends to decrease in all cases, which indicates that the interface is insufficiently discretized.
Summarizing, the cases V2 and V4 show poor accuracy compared to V1 and V3, which are discussed in Section~\ref{sec:parameterScaledCSFexample}.
Since V2 and V4 involve harmonic mean interpolation \eqref{eq:parameter_transition_harmonic}, the profile of the effective volume-specific heat capacity $\volumetricCapacityEff$ is not centered around the interface midplane, as can be seen in the top right panel of Fig.~\ref{fig:interface_heat_source_by_capacity_asymdelta}, which may contribute to the requirement of a high mesh resolution.

\end{appendices}

\FloatBarrier

\bibliography{ms}

\end{document}